\documentclass[onecolumn,superscriptaddress,amsmath,amssymb,floats,notitlepage]{revtex4-1}
\usepackage[latin9]{inputenc}
\usepackage{mathtools}
\usepackage{amssymb}
\usepackage{amsmath}
\usepackage{physics}
\usepackage{comment}
\usepackage{amsfonts}
\usepackage{graphicx}
\usepackage{wrapfig}
\usepackage{framed} %to frame figures
\usepackage{enumitem}
\usepackage{tabu}

\usepackage{tikz}
\usetikzlibrary{positioning,arrows}
\usetikzlibrary{patterns}
\usetikzlibrary{decorations.markings}
\usetikzlibrary{decorations.pathmorphing}
\usepackage{amsmath}
\usepackage{subcaption} %for subfigures

\numberwithin{equation}{section}

\newcommand{\beq}{\begin{equation}}
\newcommand{\eeq}{\end{equation}}
\newcommand{\bea}{\begin{eqnarray}}
\newcommand{\eea}{\end{eqnarray}}

\newcommand{\om}{\omega_m}
\newcommand{\Om}{\Omega_m}

\DeclareMathOperator{\sgn}{sgn}
\newcommand{\bOm}{\overline{\Omega}_m}

\newcommand{\bmuc}{\bar{\mu}_c}
\newcommand{\omd}{\omega_D}
\newcommand{\dvertex}{D^\mathrm{vertex}}
\newcommand{\dex}{D^\mathrm{exchange}}

\newsavebox{\measurebox}

\tikzset{middlearrow/.style={
        decoration={markings,
            mark= at position 0.57 with {\arrow[line width=1mm]{#1}} ,
        },
        postaction={decorate}
    }
}

\tikzset{snake it/.style={decorate, decoration=snake}}

\tikzset{
particle/.style={thick,draw=black, postaction={decorate},
    decoration={markings,mark=at position .57 with {\arrow[line width=1mm]{latex}}}},
boson/.style={decorate, draw=black,
    decoration={snake}},
dboson/.style={line width = 0.45mm, double,decorate, draw=black,
    decoration={snake}}
 }

\begin{document}
\title{Kohn-Luttinger Correction to $T_c$ in a Phonon Superconductor}
\author{Dan Phan}
\affiliation{School of Physics and Astronomy, University of Minnesota, Minneapolis, Minnesota 55455, USA}
\author{Andrey V. Chubukov}
\affiliation{School of Physics and Astronomy, University of Minnesota, Minneapolis, Minnesota 55455, USA}

\date{\today}

\begin{abstract}
Weak coupling theory predicts the critical temperature of a phonon superconductor to be $T_c = 1.13 e^{-3/2} \omega_D e^{-{1/\lambda}}$, where $\omega_D$ is the Debye frequency, $\lambda$ is the dimensionless electron-phonon coupling constant, and the factor $e^{-3/2}$ comes from fermionic self-energy and frequency dependence of the interaction. Other corrections are small either in $\omega_D/E_F$, by Migdal's theorem, or in $\lambda$. However, this formula assumes that $\omega_D \ll E_F$, where $E_F$ is the Fermi energy. We obtain $T_c$  in the dilute regime, when the Fermi energy is smaller than $\omega_D$.  We argue that in this situation Migdal's theorem is no longer valid, and Kohn-Luttinger-type corrections to the pairing interaction must be included to obtain the correct prefactor for $T_c$.
\end{abstract}

\maketitle

\begin{table}[]
    \centering
    \begin{tabular}{c|c|c|c}
    \hline
    \hline
    &$T_c$ without corrections&$T_c$ including corrections & $\mu(T_c)$  \\
    \hline
    $\omd \ll E_F$  &$1.13\omd \exp(-1/\lambda)$ &$0.25\omd \exp(-1/\lambda)$ & $E_F$\\
    $E_0 \ll E_F \ll \omd$ &$1.13 \sqrt{E_F E_0}$ &   $0.12\sqrt{E_F E_0}$ & $E_F$\\
    $E_F \ll E_0$ & $E_0/\log(E_0/E_F)$ & $4.48E_0/\log(E_0/E_F)$ & $-4.48E_0$ \\
    \hline
    \hline
    \end{tabular}
    \caption{The summary of the analytic results of this paper. The values of corrected $T_c$ include the contributions  of the self-energy, the frequency-dependence of the pairing vertex, and the dressing of the interaction.}
    \label{tab:analytic_results}
\end{table}

\section{Introduction}

This paper is devoted to the calculation of superconducting $T_c$ with the exact prefactor for phonon-mediated superconductivity in quasi-2D systems  at weak coupling, in the small density limit, when the Debye frequency is larger than the Fermi energy.

The BCS theory of phonon-mediated superconductivity~\cite{bardeen1957theory} predicts the value of superconducting $T_c=1.13\omd \exp(-1/\lambda)$. The  derivation of this formula uses three approximations. First, the frequency-dependent attraction, mediated by an  Einstein phonon with frequency $\omega_D$, is replaced by a constant within a shell of width $\omd$ around the Fermi surface. Second, a weak coupling  is assumed (dimensionless $\lambda \ll 1$) and  all corrections of $\mathcal{O}(\lambda)$ are neglected.  Third, $\omega_D$ is assumed to be much smaller than $E_F$, where $E_F$ is the Fermi energy, and all corrections small in $\omega_D/E_F$ are neglected as well.

Subsequent studies have found that  $\mathcal{O}(\lambda)$ corrections to the exponent in the BCS formula for $T_c$ actually cannot be neglected because they change the prefactor in $T_c$ by a factor $\mathcal{O}(1)$.
%AC
These corrections were studied in detail in the limit $E_F \gg \omega_D$ for electron-phonon interaction~\cite{prefac1,prefac2,prefac3,prefac4,prefac5,Chubukov2016,Marsiglio18} and for a more general case of arbitrary non-critical 
bosonic propagator~\cite{combescot1990critical}.  The corrections were argued to originate from the fermionic self-energy and the frequency dependence of the actual phonon-mediated interaction $V(\om,\om') \propto \omd^2/((\om-\om')^2+\omd^2)$.  The self-energy  $\Sigma (\om) = i \lambda \om$ changes $1/\lambda$ in the exponent to $(1+\lambda)/\lambda = 1/\lambda +1$, which changes the prefactor for $T_c$ by $e^{-1}$.  The frequency dependence of the interaction additionally changes $1/\lambda$
to $ (1 + \lambda/2)/\lambda = 1/\lambda +1/2$, i.e., changes the prefactor by $e^{-1/2}$. The full prefactor of $T_c$ is then $e^{-3/2}$, i.e., with these corrections $T_c=1.13 e^{-3/2}\omd \exp(-1/\lambda) = 0.252 \omd \exp(-1/\lambda)$.  Vertex corrections,  which give rise to Kohn-Luttinger (KL)-type renormalization of the pairing vertex, also change the argument of the exponent by $\mathcal{O}(\lambda)$. However, in the adiabatic regime where $E_F \gg \omega_D$ these corrections are smaller by $\mathcal{O}(\omega_D/E_F)$ by Migdal's theorem and can be safely neglected.

The goal of this work is to obtain expressions for $T_c$ with accurate prefactors in the situation when the coupling is still weak, but the density of carriers is sufficiently low such that $E_F < \omega_D$.  Superconductivity in this limit has attracted high interest in recent years chiefly due to advances in experimental studies of SrTiO$_3$,  where superconductivity is  present at carrier densities as low as $n\sim 10^{18}$ cm$^{-3}$~\cite{Schooley64,Schooley65,Lin14}, and in other  low-density materials, like Pb$_{1-x}$Tl$_x$Te~~\cite{Chernik1981}, half-Heusler compounds~\cite{Nakajima2015}, and single-crystal Bi~\cite{Prakashaaf8227}. A full analysis of superconductivity in these systems requires one to analyze the combined effect of phonon-mediated attraction and electron-electron repulsion~
 \cite{Gurevich1962,Schooley64,Schooley65,Chernik1981,Takada1980,Ikeda1992,Grimaldi1995,Mahan,Lin14, Nakajima2015,Prakashaaf8227,Edge2015,Ruhman2016,Gorkov2016,Gorkov2017,Ruhman2017, DHLee2015,Rademaker2016,Zhou2016,Zhou2017,Trevisan2018,Savary2017,Rowley2018,Lonzarich2018,Woelfle2018,Sadovskii2018,Sadovskii2018b,Aperis2018,Aperis2018b,grabowski1984superconductivity}. In this work, we consider only the attractive part of the interaction and explicitly compute  $T_c$ in the low-density limit.  We hope our results can be used as input for future calculations of $T_c$ which include electron-electron interactions.

The limit $E_F \ll \omega_D$ is often associated with Bose-Einstein condensation (BEC) behavior, in which fermions form bound pairs at a pairing instability temperature $T_\mathrm{ins}$, which then condense at a smaller $T_c$. However, in 3D, BEC behavior only holds at strong coupling, since there is a threshold on bound state formation. In our study we consider pairing in a quasi-2D system where the crossover from BCS to BEC behavior already holds at weak coupling and can be analyzed in a controllable way. We will obtain the pairing instability temperature  at weak coupling as a function of the two scales: $E_F$ and $E_0 = \omega_D e^{-2/\lambda}$, which denote the Fermi energy and
%DP half
the bound-state energy of two fermions in vacuum respectively(note that since we are working at weak coupling, $E_0 \ll \omd$.) For notational convenience, we label this temperature $T_c$ with the understanding that this is the onset temperature for the pairing; the actual superconducting $T_c$ is somewhat smaller due to the destructive effect from phase fluctuations.~\cite{Randeria,SadeMelo1993,Chubukov2016,pokrovsky1979properties}.

Our key results are summarized in Table  \ref{tab:analytic_results} and Fig. \ref{fig:Tc}. We obtained expressions for $T_c$ with accurate prefactors
 in three regimes: $E_F \gg \omd$, $\omd \gg E_F \gg E_0$, and $E_0 \gg E_F$.  In each regime,  $\mathcal{O}(\lambda)$ corrections to the exponent in the weak-coupling formula for $T_c$ give rise to $\mathcal{O}(1)$ numerical factors.
  At $E_F \gg \omd$, these corrections come from fermionic self-energy and from frequency dependence of the interaction, while KL corrections are small in $\omd/E_F$ and can be neglected.  In the other two regimes KL corrections are relevant and must be included to get right prefactor for $T_c$.  In particular, deep in the anti-adiabatic regime, when $E_F \ll E_0 \ll \omega_D$, KL corrections increase the value of $T_c$.
       We also compute $T_c$ numerically for values of $E_F$ across these regimes and find good agreement between numerical and analytic results.

That KL corrections to the pairing interaction are relevant at small $E_F$ is not obvious, since these corrections come from the particle-hole channel. At low carrier density, i.e., at small enough $E_F$,  the value of $\mu (T_c)$ is negative. In this situation a particle-hole bubble, taken alone, vanishes because at $\mu (T_c) <0$ the poles in the two Green's  functions in the bubble are in the same half-plane of complex frequency. If the pairing interaction is frequency independent, then all KL-type corrections to the pairing interaction (which here are proportional to particle-hole bubbles) therefore vanish~ \cite{Chubukov2016,pisani2018entanglement}. However, our interaction $V_0(\om,\om') \propto \omd^2/((\om-\om')^2+\omd^2)$ is dynamical and has poles in both half-planes of frequency. The KL correction to the pairing interaction is a convolution of the two fermionic Green's functions and the dynamical interaction, which does not vanish after frequency integration, even in the limit where $E_F$ approaches 0. To be precise, this statement holds when the bandwidth is much larger than all other energy scales in the problem. For a general bandwidth $\Lambda$, the KL correction is a function of $\Lambda/\omd$ and $E_F/\omd$.  In our analysis we assume  that $\Lambda \gg \omd$. In the opposite limit where $\Lambda \ll \omd$, the interaction can be approximated by its static form, and one retrieves previous results \cite{Chubukov2016,pisani2018entanglement} that KL corrections are irrelevant (see below and Appendix C).

\begin{figure}[h]
    \centering
	\includegraphics[width = \linewidth]{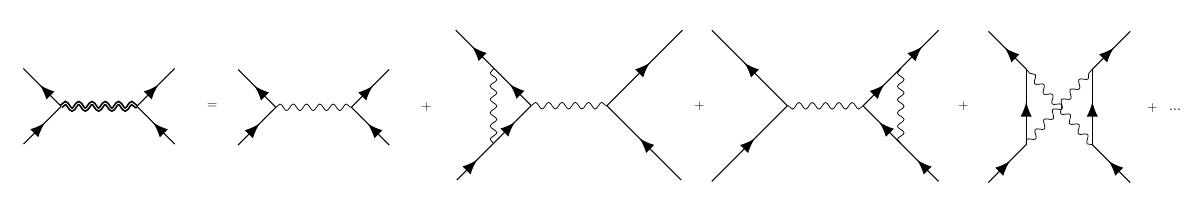}
    \caption{The diagrammatic expansion of our irreducible pairing interaction (the double wavy line).  The single wavy line is the phonon-mediated interaction $V_0(\Om) = - g \omd^2/(\Om^2 + \omd^2)$.
     We have ignored conventional screening (a diagram with an internal particle-hole bubble), as this is already included  in the bare interaction for our analysis (a screened combined Coulomb and electron-phonon interaction, the attractive part of which is  $V_0(\Om)$,  see Refs.  \cite{Gurevich1962,Takada1980,Ikeda1992,Ruhman2016,Gorkov2016,Maria2019,prokofiev,grabowski1984superconductivity}).}
    \label{fig:KL0}
\end{figure}

 We consider a model of 2D fermions with isotropic dispersion $\varepsilon(k)=k^2/2m-\mu$ and effective dynamical interaction $V_0(\om,\om') = - g \omd^2/((\om-\om')^2+\omd^2)$. The dimensionless coupling $\lambda$ is defined as $\lambda = g N_0$, where $N_0 = m/2\pi$ is the 2D density of states per spin. We follow earlier works \cite{Grimaldi1995,perali1998nonadiabatic,grabowski1984superconductivity} and assume that the RPA-type screening is already included into $V_0(\om,\om')$. Accordingly, we exclude the screening diagram from KL renormalizations. The resulting contributions to the effective interaction are shown in Fig. \ref{fig:KL0}.

%KL diagram

Our work complements several recent mean-field studies of superconductivity at low carrier density in both 3D and quasi-2D systems. The analysis of $T_c$ at $E_F \ll \omega_D$ in quasi-2D systems  up to an overall factor has been done in Refs. \cite{Randeria,SadeMelo1993,Chubukov2016} and we use the results of these works as an input for our calculations of $T_c$ with the prefactor. In Ref. \cite{Maria2019} the authors analyzed the mean-field $T_c$ in a 3D Bardeen-Pines type model with effective phonon-mediated attraction.  However, these calculations do not extend to the BEC regime. In Ref. \cite{prokofiev} the authors analyzed the combined effect of electron-electron and electron-phonon interactions at weak coupling, within the mean-field (ladder) approximation and obtained $T_c$ up to a prefactor. Our results pave the way toward extending the work in Ref. ~\cite{prokofiev} to obtain $T_c$ with the accurate prefactor. Refs. \cite{Ruhman2016,Ruhman2017} computed $T_c$ for a model with  electron-electron and electron-phonon interactions within the Eliashberg formalism. This formalism includes self-energy corrections and corrections due to the frequency dependence of the interaction, but neglects KL renormalization of the pairing interaction in the particle-hole channel. Several other works also analysed superconductivity at low carrier density assuming the system is close to a ferroelectric quantum-critical point \cite{Rowley2018, Edge2015, chandra2017prospects}. Here again, we argue that KL renormalizations must be included to obtain $T_c$ with the exact prefactor.

The paper is organized as follows. In the next section we briefly review mean-field calculations of $T_c$ up to a prefactor at $E_F \gg \omd$, $\omd \gg E_F \gg E_0$, and $E_0 \gg E_F$. In Section III we compute $\mathcal{O}(1)$ corrections to $T_c$ from fermionic self-energy and the frequency dependence of the interaction in the three ranges of $E_F$, and then discuss KL corrections to the pairing interaction. We then combine all $\mathcal{O}(1)$ corrections and present the exact results for $T_c$ in the three ranges of $E_F$. In Sec IV we present the results of our numerical calculations of $T_c$. In Section V we present our conclusions. In Appendix A, we discuss in detail the calculations of the KL corrections in the three regimes $E_F \gg \omega_D, E_0 \ll E_F \ll \omega_D$, and $E_F \ll E_0$. In Appendix B, we discuss numerical calculations of $T_c$ for a given $E_F$. Finally, in Appendix C, we discuss how KL corrections get modified for a finite fermionic bandwidth.

\section{$T_c$ to logarithmical accuracy}\label{sec:logacc}

In this section, we briefly review the derivation of $T_c$ to logarithmical accuracy (i.e., at weak coupling, up to an overall prefactor). We will find that there are 3 different expressions for $T_c$ for $E_F \gg \omd$, $\omd \gg E_F \gg E_0$, and $E_0 \gg E_F$.

%pairing vertex
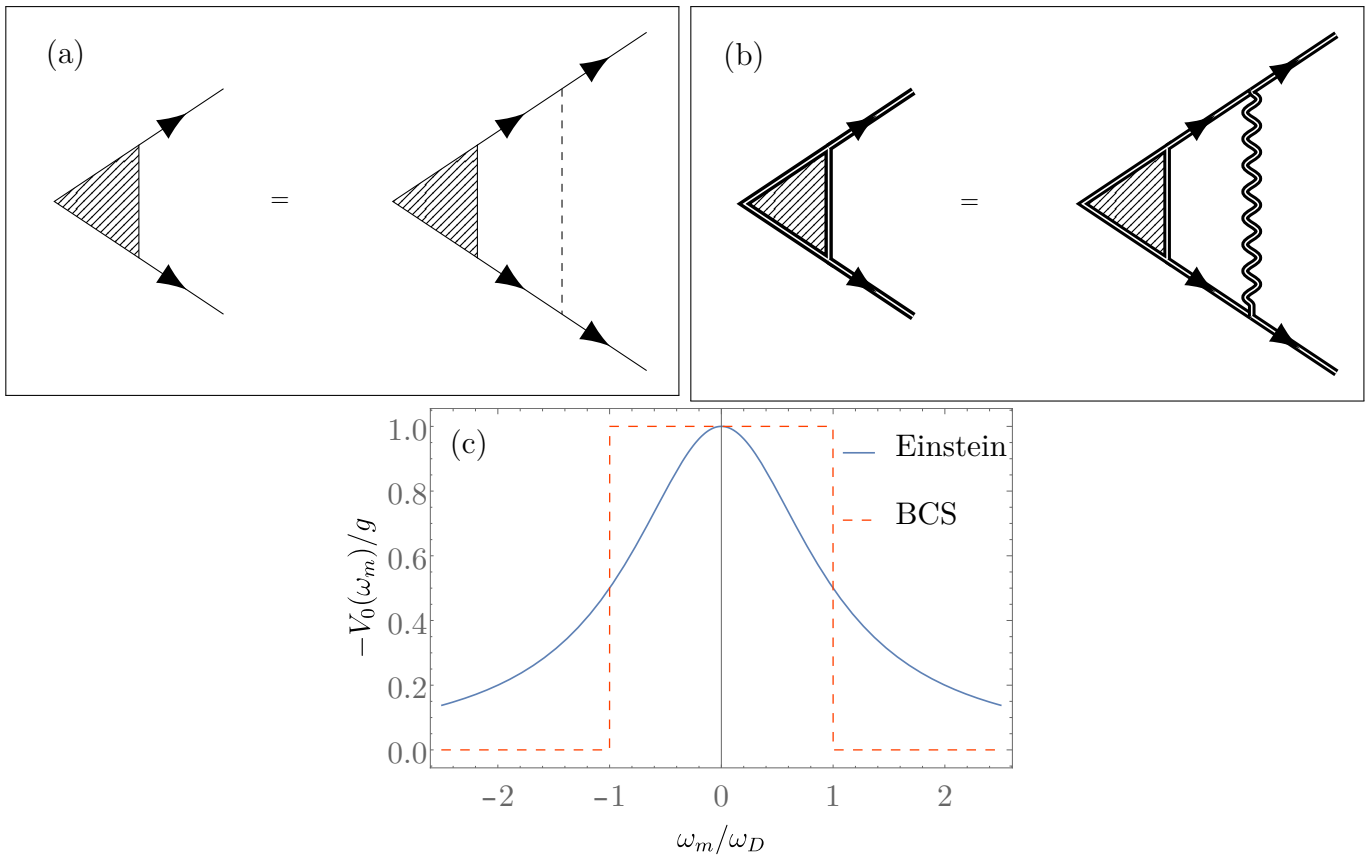
\begin{figure*}[t!]
    \centering
    \begin{subfigure}[t]{0.5\textwidth}
	\centering
	\begin{framed}
	\begin{tikzpicture}[scale=1.5]
		\draw[pattern=north east lines] (0,0) -- (.75,-.5)--(.75,.5)--cycle;
		\draw[middlearrow={latex}] (0.75,.5)--(1.5,1.);
		\draw[middlearrow={latex}] (0.75,-.5)--(1.5,-1.);
		
		\draw (2,0) node {=};
		
		\draw[pattern=north east lines] (3,0) -- (3.75,-.5)--(3.75,.5)--cycle;
		\draw[middlearrow={latex}] (3.75,.5)--(4.5,1.);
		\draw[middlearrow={latex}] (3.75,-.5)--(4.5,-1.);
		\draw[dashed] (4.5,1)--(4.5,-1);
		\draw[middlearrow={latex}] (4.5,1)--(5.25,1.5);
		\draw[middlearrow={latex}] (4.5,-1)--(5.25,-1.5);
		\draw (0.1,1.3) node {\large(a)};
	\end{tikzpicture}
    \end{framed}
    \end{subfigure}%
    ~
    \begin{subfigure}[t]{0.5\textwidth}
        \begin{framed}
        \centering
		\begin{tikzpicture}[scale=1.5]
		\draw[double, line width = 0.45mm,pattern=north east lines] (0,0) -- (.75,-.5)--(.75,.5)--cycle;
		\draw[double,line width = 0.45mm,middlearrow={latex}] (0.75,.5)--(1.5,1.);
		\draw[double, line width=0.45mm,middlearrow={latex}] (0.75,-.5)--(1.5,-1.);

		\draw (2,0) node {=};

		\draw[double, line width = 0.4mm,pattern=north east lines] (3,0) -- (3.75,-.5)--(3.75,.5)--cycle;
		\draw[double,line width=0.4mm,middlearrow={latex}] (3.75,.5)--(4.5,1.);
		\draw[double,line width=0.4mm,middlearrow={latex}] (3.75,-.5)--(4.5,-1.);
		\draw[double,line width=0.4mm, snake it] (4.5,1)--(4.5,-1);
		\draw[double,line width=0.4mm,middlearrow={latex}] (4.5,1)--(5.25,1.5);
		\draw[double, line width=0.4mm,middlearrow={latex}] (4.5,-1)--(5.25,-1.5);
		\draw (0.,1.3) node {\large(b)};
		\end{tikzpicture}
		\end{framed}
    \end{subfigure}
    \vspace{1mm}
    \begin{subfigure}{0.5\linewidth}
    \centering
    \begin{tikzpicture}
    \draw (0,0) node[inner sep=0]{\includegraphics[width = \linewidth]{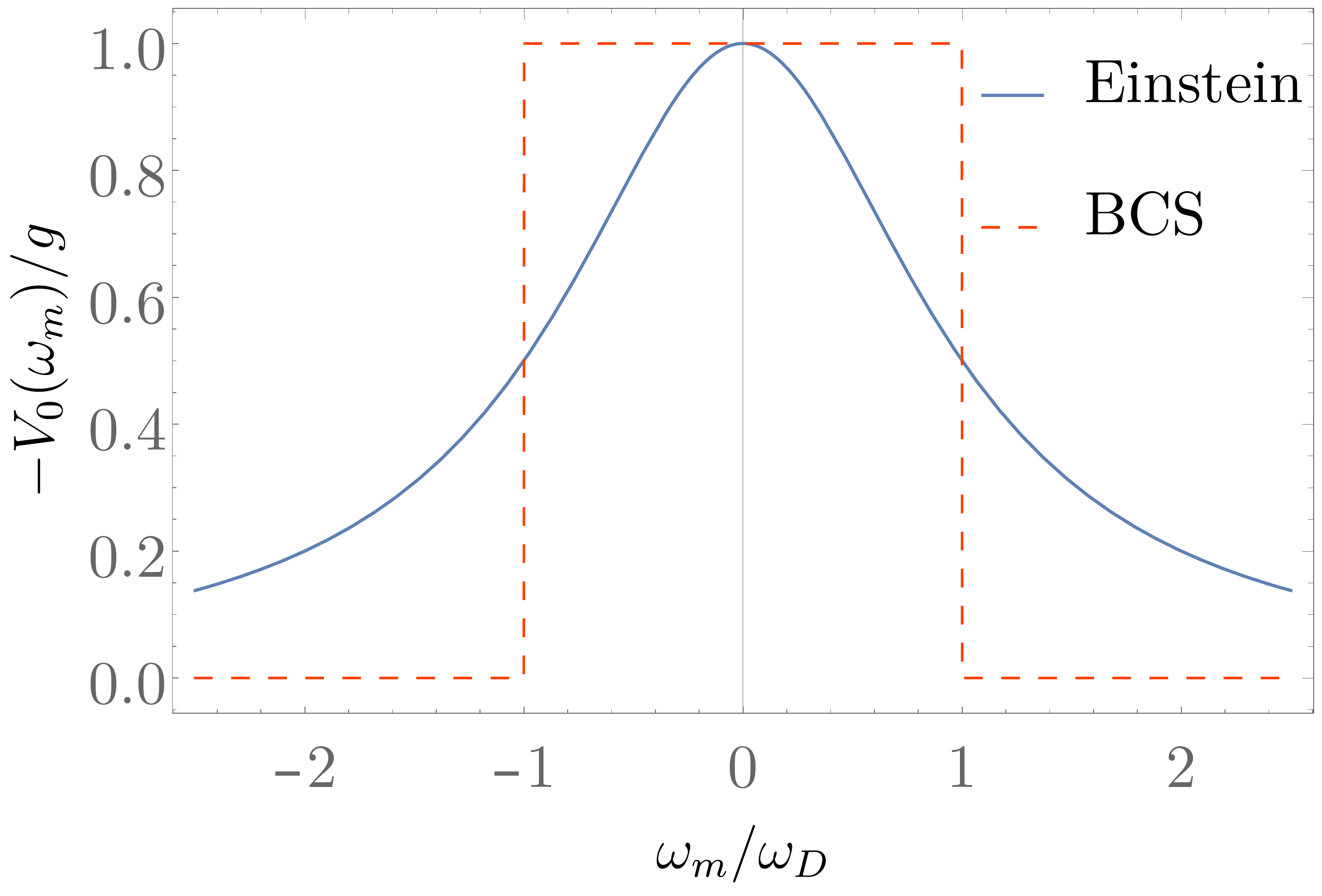}};
    \draw (-2.8,2.5) node{\large(c)};
    \end{tikzpicture}
    \end{subfigure}
    \caption{(a) The equation for the pairing vertex shown diagrammatically, within the BCS approximation. The fermionic Green's functions are bare, and the pairing interaction(dashed line) is treated as a step function. (b) The full equation for the pairing vertex. The Green's functions are fully dressed, and the interaction not only has the correct frequency dependence, but is dressed by Kohn-Luttinger contributions. (c) The Einstein and BCS approximation to the interaction potential, in units of $-g$.}
    \label{fig:pairingeqn}
\end{figure*}

%\caption{The diagrammatic equation for the pairing vertex $\Phi (\omega_m,k)$ and the form of $V(\Omega_m)$.  Panel a: BCS equation for momentum-independent $\Phi (\omega_m)$.
    %The fermionic Green's functions are bare, and the pairing interaction is treated as a step function
    %(dashed line in panel c).  Panel b: the full equation for the pairing vertex. The Green's functions are fully dressed, and the interaction is taken from Fig. 1, in which the bare interaction (a single wavy line in Fig. 1) is frequency dependent $V_0(\Om)$,
     %  shown by a solid line in panel c.  Panel c: the frequency dependence of the bare interaction.}

To obtain $T_c$ to logarithmical accuracy, we begin with the BCS equation for the pairing vertex, shown in Figure \ref{fig:pairingeqn}a. Here, the phonon-mediated interaction is approximated by a step function $V_0(\Om) = -g \Theta(\omd-\abs{\Om})$. The equation for the pairing vertex therefore becomes
\begin{equation} \label{eq:bcspairing_1}
    \Phi (\omega_m)  =  g T_c \sum_{\Om} \int \frac{d^2q}{(2\pi)^2}G(q)G(-q)\Theta(\omd-\abs{\Om - \om}) \Phi (\Om).
\end{equation}
 In addition, the particle number is fixed by
\begin{equation}\label{eq:chempot1}
    n = 2T_c \sum_{\om}\int \frac{d^2p}{(2\pi)^2}G(p).
\end{equation}
 The Green's functions in both formulas are undressed. The conventional approximation, valid to logarithmical accuracy (i.e., to leading order in $\lambda$),
 is to take $\Phi (\omega_m)$ as independent of $\omega_m$ for $|\omega_m| \ll \omd$ and ignore complications at $|\omega_m| \sim \omd$.  Setting $\Phi (\omega_m) = \Phi$ and cancelling it in Eqn. (\ref{eq:bcspairing_1}) we obtain
 \begin{equation} \label{eq:bcspairing}
    1  =  g T_c \sum_{\Om} \int \frac{d^2q}{(2\pi)^2}G(q)G(-q)\Theta(\omd-\abs{\Om - \omega_m})
\end{equation}
 The solution of (\ref{eq:bcspairing}) and (\ref{eq:chempot1}) gives $T_c$ and $\mu (T_c)$, which for notational convenience we label $\mu_c$.
 Integrating over momentum in Eq. (\ref{eq:bcspairing}) and using $n = 2N_0 E_F$, we obtain
\begin{equation} \label{eq:crude}
    1 =  \lambda T_c \sum_{\abs{\Om}<\omd} \frac{1}{\abs{\Om}}(\frac{\pi}{2}+\arctan(\frac{\mu_c}{\abs{\Om}})),
\end{equation}
and
\begin{equation} \label{eq:chempot2}
    \mu_c=T_c \log(\exp(E_F/T_c)-1).
\end{equation}
where $\lambda = g N_0$. Below we present the solution of Eqs. (\ref{eq:crude}) and (\ref{eq:chempot2}) in three ranges of values for $E_F$. As we will see, $T_c \ll \omd$ for all $E_F$.

\subsection{$E_F \gg \omd$}

In the range where $E_F \gg \omd$, we clearly have $E_F \gg T_c$. Applying this to the formula for the chemical potential, we find $\mu_c \approx E_F$. Hence, $\mu_c/\abs{\Om}>E_F/\omd \gg 1$ for all $\abs{\Om} < \omd$, and we can safely approximate  $\arctan(\mu_c/\abs{\Om})$ by $\pi/2$. Eq. (\ref{eq:crude}) then becomes
\begin{align} \label{eq:bcs}
    1
    &= \lambda \pi T_c \sum_{\abs{\Om} < \omd} \frac{1}{\abs{\Om}}\\
    &= \lambda \log(\frac{2 e^\gamma \omd}{\pi T_c}),
\end{align}
from which we find $T_c = 1.13 \omd \exp(-1/\lambda)$, the usual BCS result. The sum was done using the Euler-Maclaurin formula, using $T_c \ll \omd$.

\subsection{$E_0 \ll E_F \ll \omd$}
As $E_F$ decreases, we enter the regime where $E_F \ll \omd$, but $E_F \gg E_0 \equiv \omd \exp(-2/\lambda)$. In this range we will assume and later verify that we still have $\mu_c \approx E_F$, and $T_c \ll E_F$. Eq. (\ref{eq:crude}) then becomes

\begin{equation}
    \frac{1}{\lambda} = \frac{1}{2} \log(\frac{2 e^\gamma \omd}{\pi T_c})+ T \sum_{|\Om| < \omd} \frac{1}{\abs{\Om}}\arctan(\frac{\mu_c}{\Om}).
\end{equation}

In the second term, it is unnecessary to consider the cutoff at $\omd$ since the sum converges at $\abs{\Om} \sim \mu_c \ll \omd$. The sum can be done using the Euler-Maclaurin formula, and the equation for $T_c$ becomes

\begin{equation}
    \frac{1}{\lambda} = \frac{1}{2} \log(\frac{2 e^\gamma \omd}{\pi T_c})+ \frac{1}{2} \log(\frac{2 e^\gamma \mu_c}{\pi T_c}).
\label{new_1}
\end{equation}

Solving for $T_c$ and using $\mu_c \approx E_F$, we find $T_c = 1.13 \sqrt{\omd E_F} \exp(-1/\lambda) = 1.13 \sqrt{E_F E_0}$. Substituting this expression for $T_c$ back into (\ref{eq:chempot2}), we verify that $\mu_c \approx E_F$.

\subsection{$E_F \ll E_0 \ll \omd$}

As we further decrease $E_F$, we enter the regime where $E_F$ is much smaller than both  $E_0$ and $\omd$. To calculate $T_c$ in this limit, we will assume and then verify that $E_F \ll T_c$ and $\abs{\mu_c}\ll \omd$. Using the first assumption, we find  $\mu_c \approx T_c \log(E_F/T_c)<0$ and $\abs{\mu_c} \gg T_c$. Combining this with the second assumption, we have $\omd \gg \abs{\mu_c}\gg T_c$. The equation for $T_c$ (Eq. \ref{eq:crude}) therefore becomes

\begin{align}
    \frac{1}{\lambda}
    &=  \frac{1}{2} \log(\frac{2 e^\gamma \omd}{\pi T_c})- T_c \sum_{\Om} \frac{1}{\abs{\Om}}\arctan(\frac{\abs{\mu_c}}{\Om})\\
    &= \frac{1}{2} \log(\frac{2 e^\gamma \omd}{\pi T_c}) - \frac{1}{2} \log(\frac{2e^\gamma \abs{\mu_c}}{\pi T_c})\\
    &= \frac{1}{2}\log(\frac{\omd}{\abs{\mu_c}}).
\end{align}
Hence
\begin{equation} \label{eq:transcendental}
    \abs{\mu_c} =
    \omega_D \exp(-2/\lambda)
    = E_0,
\end{equation}
and we see that $\abs{\mu_c} \ll \omd$, as assumed.  Using $\abs{\mu_c} = T_c \log{(T_c/E_F)}$, we find
\beq
T_c = \frac{E_0}{\log(E_0/E_F)}
\eeq
 to leading order in $\log(E_0/E_F)$. This justifies our assumption that $T_c \gg E_F$.

We emphasize that these results for $T_c$ are valid only to logarithmical accuracy, i.e., up to numerical prefactors. To get $T_c$ with correct prefactors, one must include all corrections of $\mathcal{O}(\lambda)$. This is done in the following section.

\section{ $\mathcal{O}(1)$ Corrections to $T_c$ from the self-energy, frequency dependence of the interaction, and KL renormalizations}

To illustrate the point that $\mathcal{O}(1)$ corrections to $T_c$ come from $\mathcal{O}(\lambda)$ corrections to BCS  theory, consider Eq. (\ref{eq:bcs}) with an additional term $C \lambda$.
We have
\begin{align}
    1
    &= \lambda \log{\frac{2 e^\gamma \omd}{\pi T_c}} + C \lambda\\
    &= \lambda \log{\frac{2 e^\gamma e^C\omd}{\pi T_c}}
\end{align}

Solving for $T_c$ we obtain $T_c = \frac{2 e^\gamma e^C}{\pi}\omega_D e^{-\frac{1}{\lambda}}$. We see that the exponent $e^{-1/\lambda}$ is unchanged, but the prefactor has been modified by a constant $e^C$. Hence, terms of order $\mathcal{O}(\lambda)$ will affect the prefactor for $T_c$. This reasoning applies for all values of $E_F$.

To take $\mathcal{O}(\lambda)$ corrections into account, we write down the linearized equation for the full pairing vertex $\Phi(\om,\mathbf{k})$. It is given diagrammatically
 by Fig. \ref{fig:pairingeqn} b.  In analytical form we have
\begin{align} \label{eq:pairing_vertex}
    \Phi(\om,\mathbf{k})
    &=
    -T_c \sum_{\Om} \int \frac{d^2q}{(2\pi)^2}G(\Omega_m,q) G(-\Omega_m, -q) V_\mathrm{eff} (\om,\mathbf{k};\Om,\mathbf{q})\Phi(\Om,\mathbf{q}).
\end{align}
where $G(\Omega_m, q)$ is the Green function for interacting fermions, and $V_\mathrm{eff} (\om,\mathbf{k};\Om,\mathbf{q})$ is  the irreducible dynamical interaction in the particle-particle channel, dressed by renormalizations from the particle-hole channel. Eq. (\ref{eq:pairing_vertex}) must be solved along with the equation for chemical potential (Eqn. (\ref{eq:chempot2}) with the full fermionic $G$) simultaneously for $T_c$ and $\mu_c$.
For our isotropic dispersion, the pairing problem decouples between harmonics with different angular momentum $l$. Since we are interested in $T_c$ in the $s-$wave channel, the corresponding pairing vertex is $\Phi (\omega_m,k) = \Phi (\omega_m)$.

In the previous section, we approximated $G(\Omega_m,q)$ by its bare value $G_0(\Omega_m, q)=(i\Omega_m-\varepsilon_q)^{-1}$ and the irreducible pairing interaction by a step function $V_0(\om) \rightarrow -g \Theta(\omd-\abs{\om})$.
 Accordingly, we approximated the $s-$wave pairing vertex $\Phi(\om)$ by frequency-independent $\Phi$.

To find corrections $\mathcal{O}(\lambda)$ we must go beyond these approximations in three different directions:

\begin{enumerate}
    \item We must include $\mathcal{O}(\lambda)$ renormalization of the electron Green's function, $G(\Omega_m,q)$.
    \item We must take into account the frequency dependence of the bare phonon-mediated interaction $V_(\Om)$ and solve for the
     frequency dependent $\Phi(\om)$.
    \item We must include KL corrections, which account for the difference between  $V_0(\om;\Om)$  and  $V_\mathrm{eff} (\om,\mathbf{k};\Om,\mathbf{q})$.
\end{enumerate}

We emphasize that we are only interested in $\mathcal{O}(\lambda)$ corrections to the argument in the exponent for $T_c$ - these give rise to $\mathcal{O}(1)$ renormalizations of the prefactor for $T_c$. Accordingly, we neglect  regular $\mathcal{O}(\lambda)$ corrections to $T_c$. In the following, we consider each correction individually and later add the results, which is legitimate to $\mathcal{O}(\lambda)$.

\begin{figure}
\begin{tikzpicture}
\draw[double, middlearrow={latex},line width=0.45mm] (0,0) -- (2,0);
\draw (3,0) node {=};
\draw[middlearrow={latex}] (4,0) -- (6,0);
\draw (7,0) node {+};
		\draw[middlearrow={latex}] (8,0) -- (9,0);
		\draw[middlearrow={latex}] (9,0) -- (13,0);
		\draw[middlearrow={latex}] (13,0) -- (14,0);
		\draw[ snake it] (13,0) arc (0:180:2cm);
\draw (16,0) node {+ ...};
\end{tikzpicture}
\caption{The  fermionic Green's function to first order in the interaction. The tadpole correction (not shown) is already incorporated into $\mu_c$.}
\label{fig:selfenergy}
\end{figure}
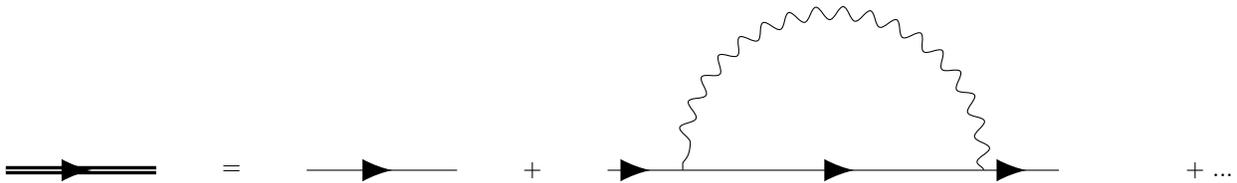

\subsection{Corrections from the fermionic self-energy}
The fermionic self-energy renormalizes the coupling $\lambda$ into $\lambda^*=\lambda/Z$, where $Z = 1 -i d\Sigma(\om)/d\om$.
The one-loop self-energy is shown in  Fig. \ref{fig:selfenergy} and is given by
\beq
 \Sigma(\omega_m,k) = T_c \sum_{m} \int \frac{d^2 q}{4\pi^2} G_0(\Omega_m,q) V_0(\Omega_m-\omega_m)
\eeq
where, we remind,  $V_0(\Omega_m) = -g \frac{\omega_D^2}{\omega_D^2+\Omega^2_m}$.
Performing the Matsubara sum, we obtain
 \beq
    \Sigma(\omega_m,k) = \Sigma (\omega_m)
    = \frac{\lambda}{2} \omd \int_{-\mu_c}^\infty d\varepsilon \big(\frac{n_F(\varepsilon)+n_B(\omd)}{\varepsilon-i\om -\omd}+\frac{1-n_F(\varepsilon)+n_B(\omd)}{\varepsilon-i\om+\omd} \big)
\eeq
Because $T_c$ is exponentially small, and $\Sigma (\omega_m)$ already contains $\lambda$ in the prefactor, the self-energy can be safely approximated by its value at $T=0$, where $n_F(\varepsilon)=\Theta(-\varepsilon)$ and $n_B(\omd)=0$.
We then end up with 2 expressions, depending on the sign of $\mu_c$.

\subsubsection{$\mu_c>0$}
 Here
\begin{equation}
    \Sigma(\om) = \Sigma(0) + \frac{\lambda\omd}{2}
    \left(\log(\frac{\omd+i\om}{\omd-i\om})-\log(\frac{\mu_c+\omd+i\om}{\mu+\omd}) \right)
    \label{new_3}
\end{equation}
Because the relevant $\om$ are of order $T_c$, i.e., exponentially smalller than $\omd$, one can expand in $\omega_m$. This gives
\beq
Z = 1+\frac{\lambda}{2} \frac{2 \mu_c + \omd}{\mu_c + \omd}
\eeq
At $E_F \gg \omd$,  $\mu_c \approx E_F \gg \omd$, and $Z = 1+\lambda$. This is a well-known result \cite{coleman2015introduction}. At $E_0 \ll E_F \ll \omd$, we have $\mu_c \approx E_F \ll \omd$, and $Z = 1 + \lambda/2$ instead.

\subsubsection{$\mu_c<0$}
 For $\mu_c < 0 $, only one of the two integrals survives. Now
\begin{equation}
    \Sigma(\om)-\Sigma(0)=\frac{\lambda\omd}{2}
    \log(\frac{\abs{\mu_c}+\omd}{\abs{\mu_c}+\omd-i\om})
\label{new_4}
\end{equation}
This  yields
\beq
Z=1+\frac{\lambda}{2} \frac{\omd}{\abs{\mu_c}+\omd}.
\eeq
Since $|\mu_c| \ll \omega_D$ for negative $\mu_c$, we have
 $Z=1+\lambda/2$.

Eqs. (\ref{new_3}) and (\ref{new_4}) can be combined into
\begin{equation}
    \Sigma(\om) = \Sigma (0) +
    i\om \frac{\lambda}{2}
    \frac{|\mu_c| + \mu_c + \omd}{\abs{\mu_c} + \omd},
\end{equation}
which holds for both positive and negative $\mu_c$.

With this, we may now derive modified expressions for $T_c$ in all three regimes of $E_F$, by simply replacing $\lambda \rightarrow \lambda^*= \lambda/Z$ in the expressions for $T_c$ in the previous section. The effect on the prefactor of $T_c$ due to the inclusion of the self-energy in all three cases is summarized in Table \ref{tab:prefactors}. We recall that $T_c \propto e^{-1/\lambda}$ when $E_F \gg E_0$ and $T_c \propto e^{-2/\lambda}$ when $E_F \lesssim E_0$.  In all cases, including the self-energy reduces $T_c$.

\subsection{Correction to $T_c$ from  the frequency dependence of $V_0(\Omega_m)$}

Next we obtain the $\mathcal{O}(1)$ correction to $T_c$ from the frequency dependence of the electron-phonon interaction $V_0(\Omega_m)$. For $E_F \gg \omd$, this has been considered in Refs. ~\cite{Marsiglio18,Wang13,prefac1,prefac2,prefac3,prefac4,prefac5}. We analyze the correction to $T_c$ in all three regions of $E_F$. We follow the computational approach used in~\cite{Wang13,Marsiglio18}.

We start with the Eq. (\ref{eq:pairing_vertex}) for the frequency-dependent pairing vertex at $T_c$, which we rewrite as
\begin{equation}
    \Phi(\om)=
    \frac{\lambda}{N_0}T_c \sum_{\Om} \int \frac{d^2q}{(2\pi)^2}\frac{1}{\Om^2+\varepsilon_q^2}\frac{\omd^2}{\omd^2+(\om-\Om)^2} \Phi(\Om)
\label{new_6}
\end{equation}
The leading, logarithmical contribution to the r.h.s. of (\ref{new_6}) comes from small internal $\Omega_m$, for which $\Phi(\Om)/(\omd^2+(\om-\Om)^2) \approx \Phi (0)/(\omd^2 + \om^2)$. Accordingly, we search for the solution of (\ref{new_6}) in the form
\begin{equation}
    \Phi(\om) =\Phi(0) \left(\frac{\omd^2}{\omd^2+\om^2}+\lambda \delta\Phi(\om)\right)
\end{equation}
We substitute this into (\ref{new_6}) and set  external $\omega_m$ to have the smallest possible value $\omega_m = \pi T_c$. Because $T_c$ is much smaller than typical $\Omega_m$ in all three regimes, we can safely neglect $\omega_m = \pi T_c$ compared to $\Omega_m$ in the r.h.s. of (\ref{new_6}). We find

\begin{equation}
    1+\lambda \delta \Phi(0)= \frac{\lambda}{N_0}T_c \sum_{\Om}\int \frac{d^2q}{(2\pi)^2}\frac{1}{\Om^2+\varepsilon_q^2}\frac{\omd^2}{\omd^2+\Om^2}\left(\frac{\omd^2}{\omd^2+\Om^2}+\lambda \delta \Phi(\Om)\right).
    \label{new_7}
\end{equation}
To the same accuracy, the last term in the right-hand side can be approximated as
\begin{equation}
    \lambda \delta \Phi(0) I
 \end{equation}
 where
 \beq
    I = \frac{\lambda}{N_0}T_c \sum_{\Om}\int \frac{d^2q}{(2\pi)^2}\frac{1}{\Om^2+\varepsilon_q^2}\frac{\omd^2}{\omd^2+\Om^2}.
\eeq
One can verify that $I = 1 + \mathcal{O}(\lambda)$. Substituting this back into (\ref{new_7}), we obtain
\begin{equation}
    \frac{\lambda}{N_0}T_c \sum_{\Om}\int \frac{d^2q}{(2\pi)^2}\frac{1}{\Om^2+\varepsilon_q^2}\left(\frac{\omd^2}{\omd^2+\Om^2}\right)^2
     = 1+\lambda \delta \Phi(0) (1-I) = 1 + \mathcal{O}(\lambda^2)
    \label{new_8}
\end{equation}
To order $\mathcal{O}(\lambda)$ we then have
\begin{equation}
    \frac{\lambda}{N_0}T_c \sum_{\Om}\int \frac{d^2q}{(2\pi)^2}\frac{1}{\Om^2+\varepsilon_q^2}\left(\frac{\omd^2}{\omd^2+\Om^2}\right)^2
     = 1.
 \label{eq:almostfullgap}
\end{equation}
Comparing this with  Eq. (\ref{eq:pairing_vertex}), we see that the effect of the frequency dependence of the pairing vertex is that $\Theta(\omd-\abs{\Om}) \rightarrow (\omd^2/(\omd^2+\Om^2))^2$. This difference becomes relevant at frequencies comparable to $\omd$.

  Integrating over momentum in (\ref{eq:almostfullgap})  we find
\begin{equation}
    1= \lambda T_c \sum_{\Om}\left(\frac{\omd^2}{\omd^2+\Om^2}\right)^2 \frac{1}{\abs{\Om}}(\frac{\pi}{2}+\arctan(\frac{\mu_c}{\abs{\Om}})).
\label{new_9}
\end{equation}
At $E_F \gg \omd$, this reduces to
\begin{equation}
    1= \lambda \pi T_c \sum_{\Om}\left(\frac{\omd^2}{\omd^2+\Om^2}\right)^2 \frac{1}{\abs{\Om}} = \log{\frac{2e^{\gamma}\omd}{\pi T_c \sqrt{e}}}
\end{equation}
Comparing with $T_c$ in the previous section, we see that $T_c$ is reduced by $\sqrt{e}$.

For $E_F \ll \omd$, recall that the second term on the r.h.s. of Eq. \ref{new_9} converges at $\abs{\Om} \sim \abs{\mu_c} \ll \omd$. Therefore, the modification $\Theta(\omd-\abs{\Om}) \rightarrow (\omd^2/(\omd^2+\Om^2)^2)^2$ has no effect on this sum. In this situation, the renormalization factor for $T_c$ comes only from the $\pi/2$ term, and equals to $1/e^{1/4}$ instead of $1/\sqrt{e}$.  For smaller Fermi energies ($E_F \lesssim E_0$), the only contribution to the renormalization is again from the $\pi/2$ term in the r.h.s. of Eq. (\ref{new_9}). However, since $T_c \propto e^{-2/\lambda}$, the renormalization factor is $(1/e^{1/4})^2 = 1/\sqrt{e}$.

\subsection{KL renormalization of the pairing interaction}\label{sec:kl}

We now take into account the first-order KL correction to the interaction. We express the dressed interaction as $V_\mathrm{eff} (\omega_m,k;\omega'_m,k')$ as $V_\mathrm{eff} (\omega_m,k;\omega'_m,k') = V_0(\om-\om')+\lambda \delta V(\omega_m,k;\omega'_m,k') + \mathcal{O}(\lambda^2)$. The point of this section is to calculate the effect of $\lambda \delta V(k,k')$ on $T_c$. For convenience, we pull out the coupling constant $g$ and express $V_\mathrm{eff} (\omega_m,k;\omega'_m,k') = -g D_\mathrm{eff}(\omega_m,k;\omega'_m,k')$, where $D_\mathrm{eff} (\omega_m,k;\omega'_m,k') = D_0(\om-\om') + \lambda \delta D (\omega_m,k;\omega'_m,k')$ is dimensionless.

The KL diagrams for $\delta D (\omega_m,k;\omega'_m,k')$ are shown in Figure $\ref{fig:KL0}$.
There are 3 first order corrections to the bare interaction. The first two describe vertex corrections, and the third is the exchange (crossing) diagram.

Before calculating $\delta D (\omega_m,k;\omega'_m,k')$ explicitly, we show how it modifies $T_c$.
For this we go back to Eq. (\ref{eq:pairing_vertex}) for the pairing vertex $\Phi(\om,\mathbf{k})$, explicitly express $V_\mathrm{eff}$ as the sum of the two terms, and neglect other $\mathcal{O}(\lambda)$ corrections, i.e., approximate $G$ by its free fermion value and approximate $V(\Omega_m)$ by a step function.
The equation for the pairing vertex then reduces to

\begin{align}
    \Phi(\om,\mathbf{k})
    &=
    \lambda \frac{T_c}{N_0} \sum_{\Om} \int \frac{d^2q}{(2\pi)^2} \frac{1}{\Om^2+\varepsilon_q^2} \left[\Theta(\abs{\Om-\om}-\omd)+\lambda \delta D (\om,\mathbf{k};\Om,\mathbf{q})\right] \Phi(\Om,\mathbf{q}).
\end{align}

Due to the factor of $(\varepsilon_q^2+\Om^2)^{-1}$, the integrand peaks at $q=k_\mu \equiv \sqrt{2m\mu_c}$ for $\mu_c>0$ and at $q=0$ for $\mu_c < 0$. In $\delta D (\om,\mathbf{k};\Om,\mathbf{q})$ we then set $\mathbf{k} = \mathbf{n}_k k_\mu \Theta(\mu_c)$ and $\mathbf{q} = \mathbf{n}_q k_\mu \Theta(\mu_c)$. Like before, we set $\om = \pi T_c$ and set the pairing vertex to be a nonzero constant for $\abs{\om} < \omd$ and 0 for $\abs{\om} > \omd$, mirroring the frequency dependence of the bare interaction. We then obtain
\beq \label{eq:fullgap}
    1 =  \lambda \frac{T_c}{N_0}\sum_{|\Om| < \omd} \int \frac{d^2q}{(2\pi)^2} \frac{1}{\Om^2+\varepsilon_q^2}
     + \lambda^2 \frac{T_c}{N_0}\sum_{|\Om| < \omd} \int \frac{d^2q}{(2\pi)^2} \frac{1}{\Om^2+\varepsilon_q^2}
    \delta D (0,\mathbf{n}_k k_\mu \Theta (\mu_c);\Om,\mathbf{n}_q k_\mu \Theta (\mu_c))
\eeq

For the last term there is a logarithmical contribution at $\Om = 0$ and $q=k_\mu$, which cancels one power of $\lambda$. Accordingly, we set $\Om=0$ in
$\delta D$. The part of the KL interaction relevant for our purposes, is therefore  $\delta D (0, \mathbf{n}_k k_\mu \Theta(\mu_c) ;0, \mathbf{n}_q k_\mu \Theta(\mu_c))$. We still need to integrate over the angle between $\mathbf{n}_k$ and $\mathbf{n}_q$
 as we are computing $T_c$ for s-wave pairing.  We therefore define $\overline{\delta D}=\int_0^{2\pi} \frac{d\theta}{2\pi} \delta D (\theta)$, where $\theta$ is the angle between $\mathbf{n}_k$ and $\mathbf{n}_q$. Using that to first order in $\lambda$,
 \beq
  \lambda \frac{T_c}{ N_0} \sum_{|\Om|<\omd} \int_0^\infty \frac{q dq}{2\pi} \frac{1}{\Om^2+\varepsilon_q^2} = 1,
 \eeq
we obtain from (\ref{eq:fullgap})
\begin{equation} \label{eq:fullpairing}
    \lambda \frac{T_c}{N_0}\sum_{|\Om| < \omd} \int \frac{d^2q}{(2\pi)^2} \frac{1}{\Om^2+\varepsilon_q^2}
    = 1 - \lambda\overline{\delta D} + \mathcal{O}(\lambda^2).
\end{equation}
We see that the KL renormalization of the interaction changes $1/\lambda$ to $1/\lambda -  \overline{\delta D }$. For $E_F \gtrsim E_0$, $T_c \propto e^{-1/\lambda}$ then acquires a factor $e^{\overline{\delta D }}$. For $E_F \lesssim E_0$, $T_c \propto e^{-2/\lambda}$, and the factor is $e^{2\overline{\delta D}}$.

The calculation of $\overline{\delta D}$ is somewhat involved and is presented in Appendix A. The results are as follows:  for $E_F \gg \omd$, $\overline{\delta D}$ is small in $\omega_D/E_F$, in agreement with Migdal's theorem.
At $E_0 \ll E_F \ll \omd$, we find $\overline{\delta D}= -3/2$, so the KL renormalization reduces $T_c$ by $e^{3/2}$. At $E_F \ll E_0$, we find $\overline{\delta D}= 3/2$. Hence, the KL renormalization increases $T_c$ by $e^3$.

The sign change of $\overline{\delta D}$ between $E_0 \ll E_F \ll \omd$ and $E_F \ll E_0$ is specific to 2D and can be understood by analytically computing $\delta D$ at $T=0$.  The sign change occurs at $E_0 = E_F$, when $\mu_c$ changes sign.
To see this, we note that each  diagram for $\delta D$ in Fig. 1 is the convolution of the interaction $V_0(\Om)$ and two Green's functions. For $E_F \sim E_0 \ll \omega_D$, the relevant internal momenta and frequencies in the Green's functions are much larger than the relevant external ones. Therefore, up to an overall factor, each KL term is given by
\beq
J = \int_{-\infty}^\infty d \Omega_m \frac{\omega^2_D}{ \Omega^2_m + \omega^2_D}  \int^\Lambda_{-\mu} \frac{d \varepsilon}{(i \Omega_m - \varepsilon_+)(i \Omega_m - \varepsilon_-)},
\label{n_1}
\eeq
 where we have introduced a cutoff $\Lambda$, and $\varepsilon_+$ and $\varepsilon_-$ are the energies for two nearly coinciding momenta. That is, the difference between relevant $\varepsilon_+$ and $\varepsilon_-$ are on the order of $\abs{\mu_c} \ll \omd$, while typical $\varepsilon_+$ and $\varepsilon_-$ are on the order of $\omd$.

The integral over $\Omega_m$ and $\varepsilon$ in (\ref{n_1}) is not singular and can be integrated in any order. Let us first integrate over $\Om$. Consider the case $\mu_c <0$. Since $\varepsilon_{+},\varepsilon_- >0$ for negative $\mu_c$, the frequency integral is entirely determined by the pole in the bosonic propagator at $\Om = i \omd$. Once this pole is taken, we can safely set $\varepsilon_+ = \varepsilon_- = \varepsilon$ and integrate over dispersion. The integrand is singularity-free, and we obtain
   \beq
  J_{\mu_c <0} = \pi \frac{\omega_D}{\omega_D + |\mu_c|}  \frac{\Lambda - |\mu_c|}{\Lambda + \omega_D}
  \label{n_2}
  \eeq

For positive  $\mu_c$ there are two contributions to $J$: $J_{\mu_c >0} = J_{1, \mu_c >0} + J_{2, \mu_c >0}$.  The contribution ($J_{1,\mu_c >0}$) again comes from the pole in the bosonic propagator. For this one can set, as before, $\varepsilon_+ = \varepsilon_- = \varepsilon$ and take the pole of $V_0(\Omega_m)$ in the frequency half-plane where there is no double pole in the fermionic propagator.  Afterwards, one can integrate over $\varepsilon$. This procedure is again free from singularities, and the result is
   \beq
  J_{1,\mu_c >0} =  \pi \frac{\omega_D}{\omega_D + \mu_c}  \frac{\Lambda  + \mu_c}{\Lambda + \omega_D} + 2\pi \frac{\Lambda \mu_c}{(\omega_D + \mu_c) (\Lambda + \omega_D)}
  \label{n_3}
  \eeq
  At $\mu_c =0$, this term coincides with the one in Eq. (\ref{n_2}).

The second contribution comes from the split poles in the fermionic propagators, from the range where $\varepsilon_+$ and $\varepsilon_-$ have opposite signs.  Because $\abs{\mu_c}$ is much smaller than $\omega_D$, the corresponding $\Om$ are small compared  to $\omd$. The term $  J_{2,\mu_c >0}$ is then, up to an overall factor, the product of the static interaction (set equal to $1$ in Eq. (\ref{n_1})) and the static particle-hole susceptibility. The latter is independent of $\mu_c$(for $\mu_c > 0$) in 2D and is equal to $-2\pi$. We hence have
   \beq
  J_{2,\mu_c >0} =  -2\pi.
  \label{n_4}
  \eeq
  Combining Eqs. (\ref{n_3}) and (\ref{n_4}), we find that near $\mu_c =0$, $J_{\mu_c >0}$ has an additional $-2\pi$ compared to $J_{\mu_c <0}$:
 \beq
  J_{\mu_c >0} =  \pi \frac{\omega_D}{\omega_D + \mu_c}  \frac{\Lambda  + \mu_c}{\Lambda + \omega_D} - 2\pi \frac{\omega_D (\Lambda + \omega_D + \mu_c)}{(\omega_D + \mu_c) (\Lambda + \omega_D)}.
  \label{n_4_1}
  \eeq

Therefore, the KL contribution to the pairing vertex, and hence, to the prefactor of $T_c$, jumps by a finite value between $E_F \gtrsim E_0$, where $\mu_c >0$ and $E_F \lesssim E_0$, where $\mu_c <0$.

This discontinuity is in fact artificial, because we computed $J$ at $T=0$, when the static particle-hole susceptibility $\chi (\mu_c)$ is discontinuous at $\mu_c=0$. At finite $T = T_c$, it is continuous, but varies rapidly in the range $|\mu_c| \leq T_c$.  Accordingly, the KL correction to the exponent is continuous, but varies rapidly around $E_F \sim E_0$. We  note in passing that  the same discontinuity between $J_{\mu_c >0}$ and $J_{\mu_c <0}$  can be obtained if one approximates $V_0(\Om)$ by a step function.

We also note that the magnitude of the KL renormalization for $\mu_c <0$ depends on the ratio $\Lambda/\omega_D$.  For $\Lambda \gg \omega_D$,  the magnitude of the KL correction is the same at positive and negative $\mu_c$, only the sign is different:
 $J_{\mu_c <0} \approx \pi, J_{\mu_c >0} \approx -\pi$. For $\Lambda \ll \omega_D$, the KL renormalization at $\mu_c <0$ becomes parametrically small: $J_{\mu_c <0} \approx \pi \Lambda/\omega_D \ll 1$.  This last result is consistent with earlier studies, which have found \cite{Chubukov2016,pisani2018entanglement} that for a static interaction the KL renormalization vanishes for $\mu_c <0$. To verify this, it is convenient to evaluate $J_{\mu_c <0}$  by integrating over $\varepsilon$ first. Doing so, one finds that typical frequencies are of order $\Lambda$. Hence, for  $\Lambda \ll \omega_D$ the interaction term
      $\omega^2_D/(\omega^2_D + \Omega^2_m)$ can be treated as static.

\begin{figure}
    \centering
    \begin{subfigure}[t]{0.3 \linewidth}
    \centering
    \includegraphics[width=\linewidth]{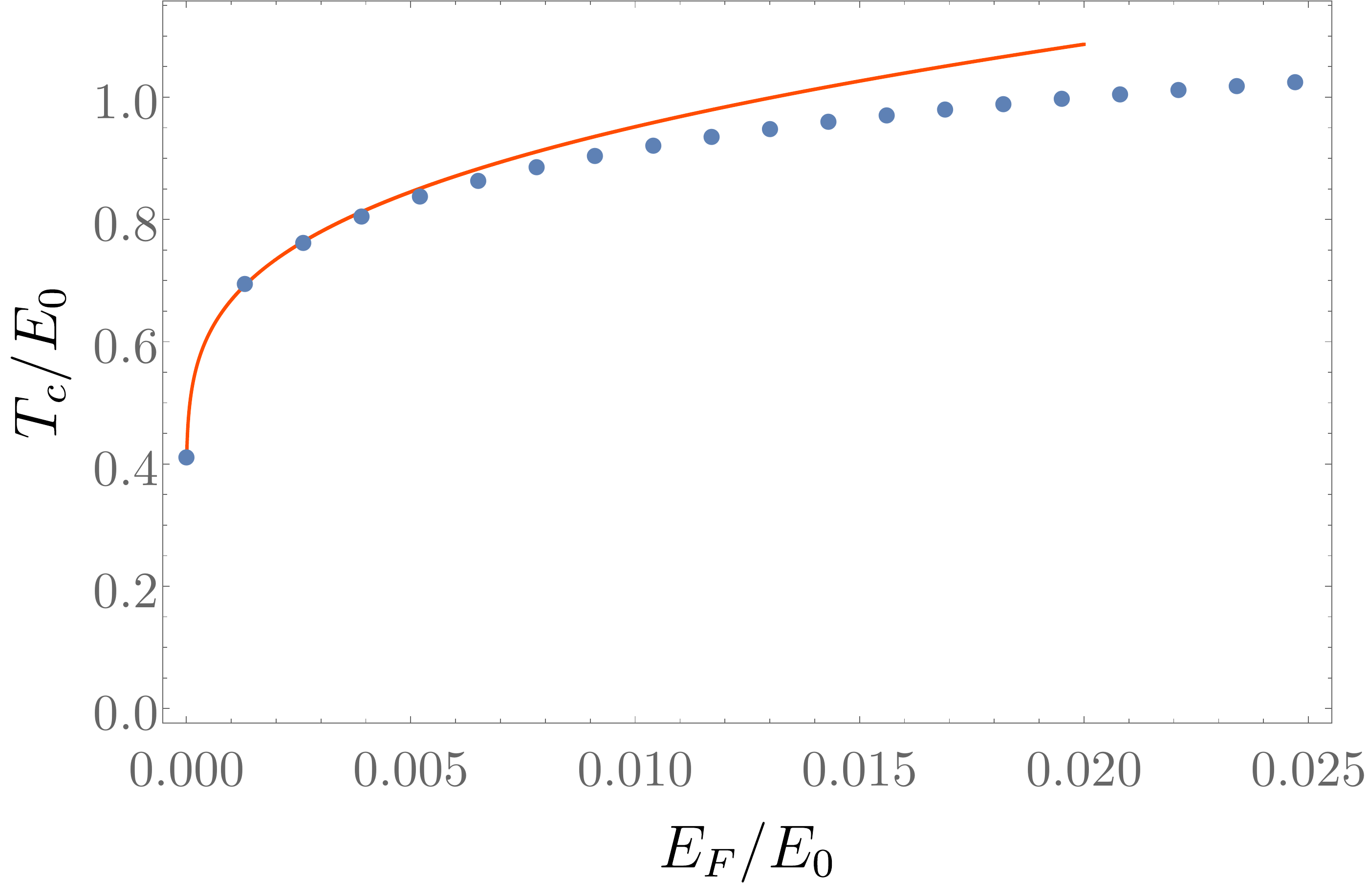}
    \end{subfigure}
    \begin{subfigure}[t]{0.28 \linewidth}
    \centering
    \includegraphics[width=\linewidth]{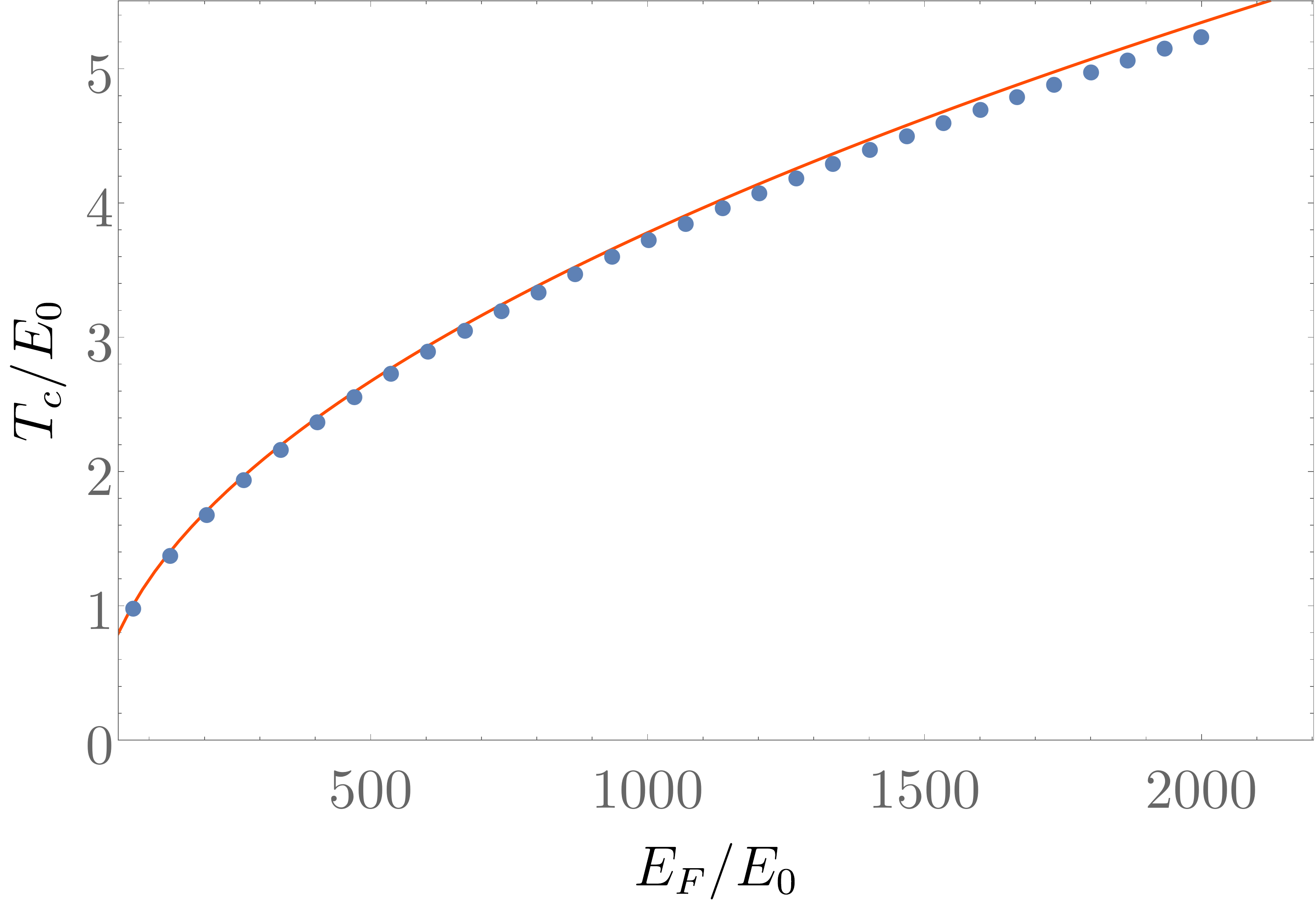}
    \end{subfigure}
   \begin{subfigure}[t]{0.33 \linewidth}
    \centering
    \includegraphics[width=\linewidth]{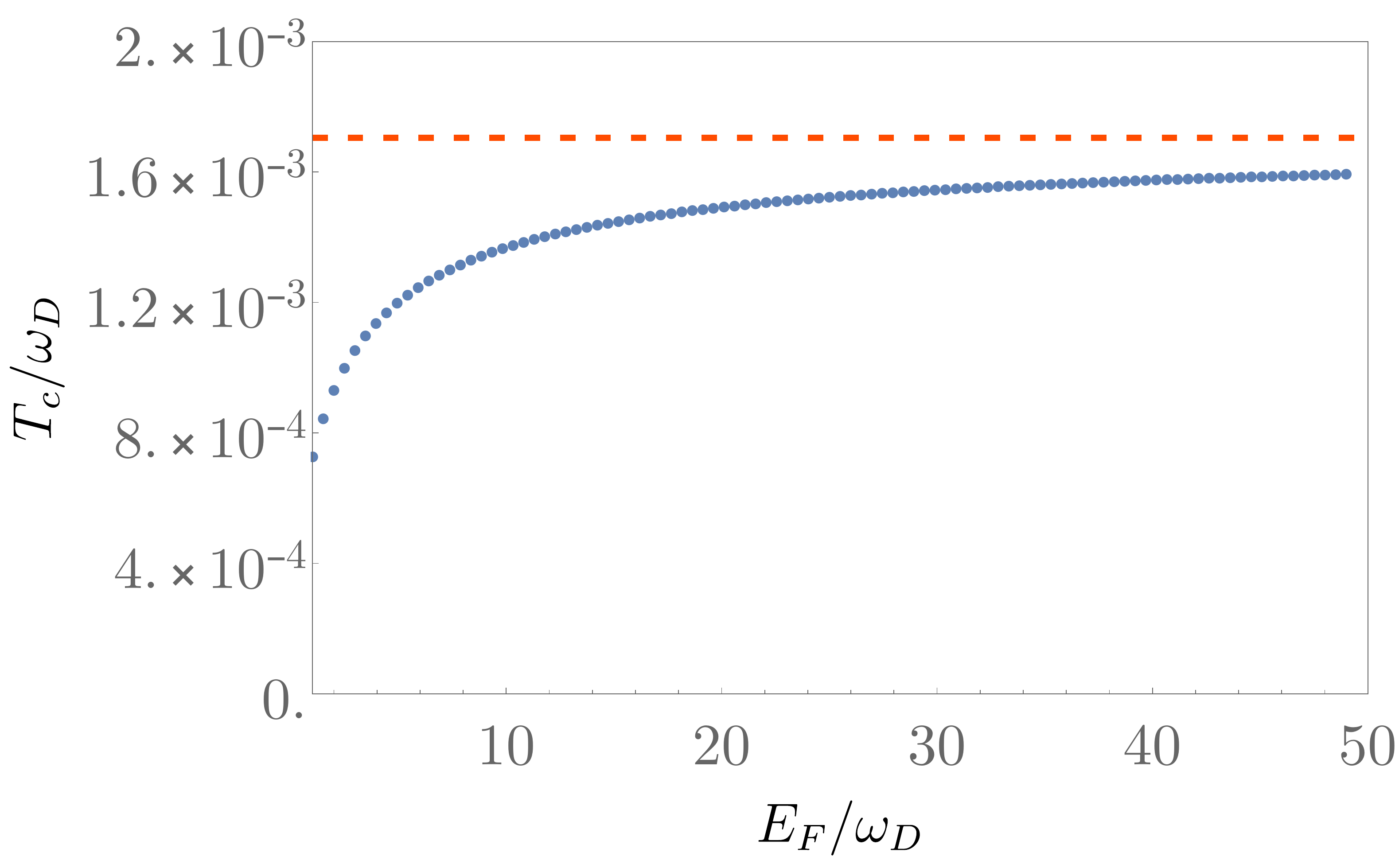}
    \end{subfigure}
\caption{$T_c$ as a function of $E_F$ in the three regimes $E_F \gg \omega_D$, $\omega_D \gg E_F \gg E_0$, and $E_0 \gg E_F$. The orange curves are our analytic expressions for $T_c$, derived in each region. At $E_F \ll E_0$ we included into the analytical expression the leading corrections to $T_c$ of order $\mathcal{O}(\log\log(E_0/E_F)/\log(E_0/E_F))$. Our numerical results for $T_c$ were found by self-consistently solving for $T_c$ and $\mu_c$ as a function of $E_F$.  We used $\lambda =0.2$, whereby $E_0 \approx 0.45 \times 10^{-4} \omega_D$. The limiting value of $T_c$ at large $E_F/\omd$ is $T_c = 0.252 \omega_D e^{-1/\lambda} \approx 1.7 \times 10^{-3} \omega_D$ (the dashed line in the last panel).}
\label{fig:Tc}
\end{figure}

\subsection{Total $\mathcal{O}(\lambda)$ corrections to the exponent, and the renormalization of $T_c$ }\label{sec:total}

We now combine the renormalizations from self-energy, frequency dependence of the interaction, and KL renormalization. To first order in $\lambda$, the numerical prefactor for $T_c$ is the product of the renormalizations from these three sources. Our analytical results for these prefactors are shown in Table \ref{tab:prefactors}.

\begin{table}[]
    \centering
    \begin{tabular}{c|c|c|c|c}
    & $\Sigma(\om)$ & $\Phi(\om)$ & KL& Total  \\
    \hline
    $\omd \ll E_F$           & $e^{-1}$   & $e^{-1/2}$  & 1   &  $e^{-3/2}$    \\
    $E_0 \ll E_F \ll \omd$   & $e^{-1/2}$ & $e^{-1/4}$  & $e^{-3/2}$  &  $e^{-9/4}$  \\
    $E_F \ll E_0$            & $e^{-1}$   & $e^{-1/2}$  & $e^{3}$        &  $e^{3/2}$ \\
    \hline
    \hline
    \end{tabular}
    \caption{The summary of the analytic results of this paper regarding modifications to the prefactor of $T_c$ from all corrections of $\mathcal{O}(\lambda)$. Also listed is the total correction to the prefactor of $T_c$, obtained by multiplying the factors from each contribution together.}
    \label{tab:prefactors}
\end{table}

\subsubsection{The case $E_F \gg \omd$}

Here only self-energy and frequency dependence of the interaction affect the prefactor for $T_c$. The result is
\beq
T_c =0.252\omd \exp(-1/\lambda).
 \eeq
This formula has been obtained earlier \cite{Wang13,Marsiglio18,prefac1,prefac2,prefac3,prefac4,prefac5}, and is presented here for completeness.

\subsubsection{The case $E_0\ll E_F \ll \omd$}

In this regime, we have
\begin{align}
    T_c &= \frac{2e^{\gamma}}{\pi}e^{-1/4}e^{\overline{\delta D}}\sqrt{\omd E_F}\exp(-Z/\lambda)\\
    &= 0.12 \sqrt{\omd E_F}\exp(-1/\lambda).
\end{align}

\subsubsection{The case $E_F \ll E_0$}

In this regime we have $ T_c = \abs{\mu_c} /\log{(\abs{\mu_c}/E_F)}$ and
\beq
  \abs{\mu_c} = e^{2\overline{\delta D}}e^{-1/2}\omd \exp(-2Z/\lambda) = e^{3/2}E_0 \approx 4.48 E_0
\eeq
Hence, to leading order in $\log{E_0/E_F}$, we obtain
\beq
 T_c = 4.48 \frac{E_0}{\log{(E_0/E_F)}}
\eeq

   \begin{figure}[t]
    \centering
    \includegraphics[width=0.5 \linewidth]{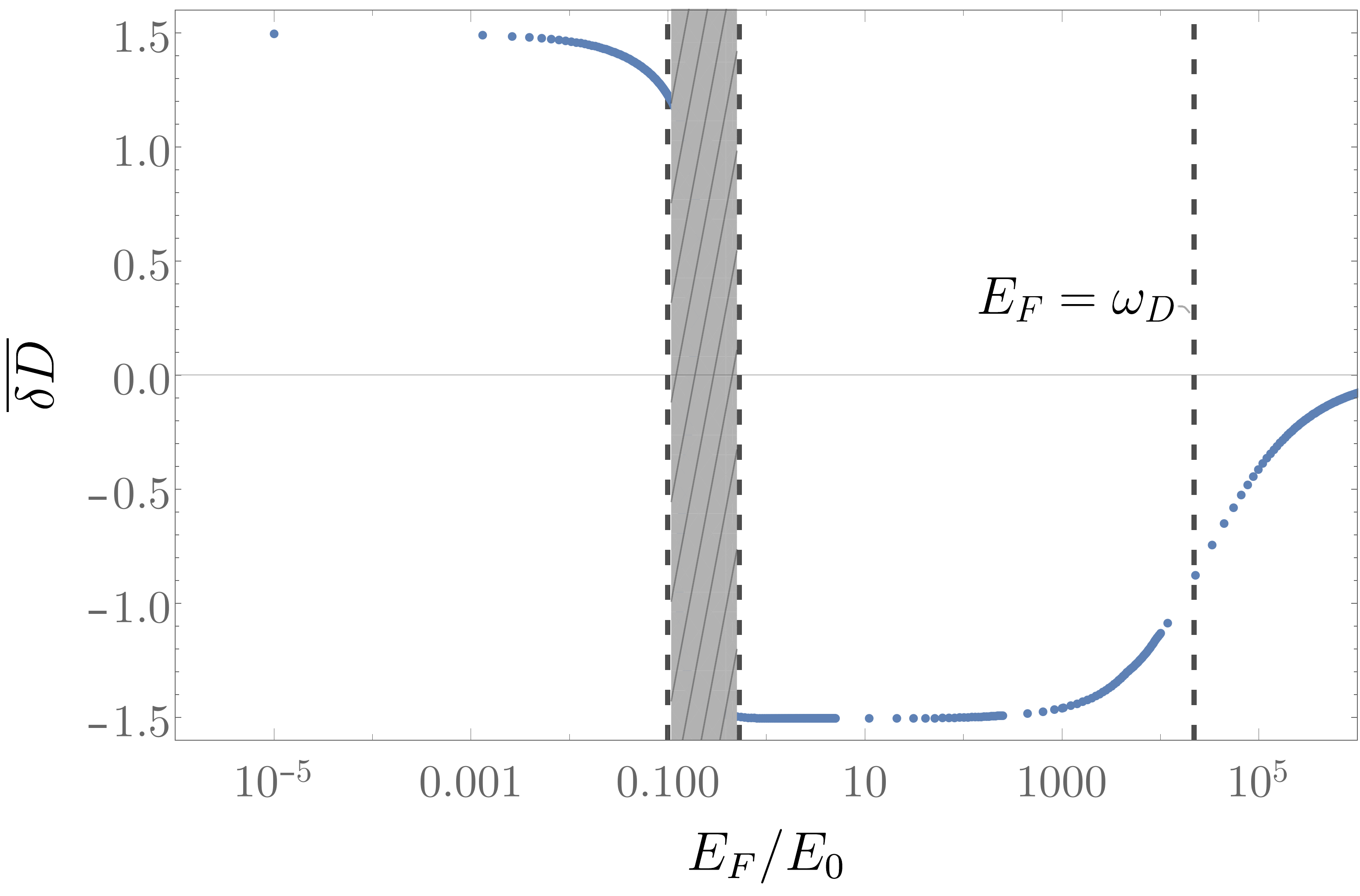}
\caption{The result of the numerical evaluation of the KL correction to $s-$wave pairing interaction, $\overline{\delta D}$, as a function of $E_F/E_0$ for a wide range of $E_F$. The KL correction is small at $E_F \gg \omega_D$, in agreement with Migdal's theorem, but becomes sizable at smaller $E_F$ and evolves from  $\overline{\delta D} \approx -1.5$ to $\overline{\delta D} \approx 1.5$ at $E_F \sim 0.17 E_0$.  The behavior near $E_F \sim 0.17 E_0$ (the shaded region in the figure) requires more
 detailed consideration. We show the behavior in this region in Fig. \ref{fig:kl}.}
\label{fig:D}
\end{figure}

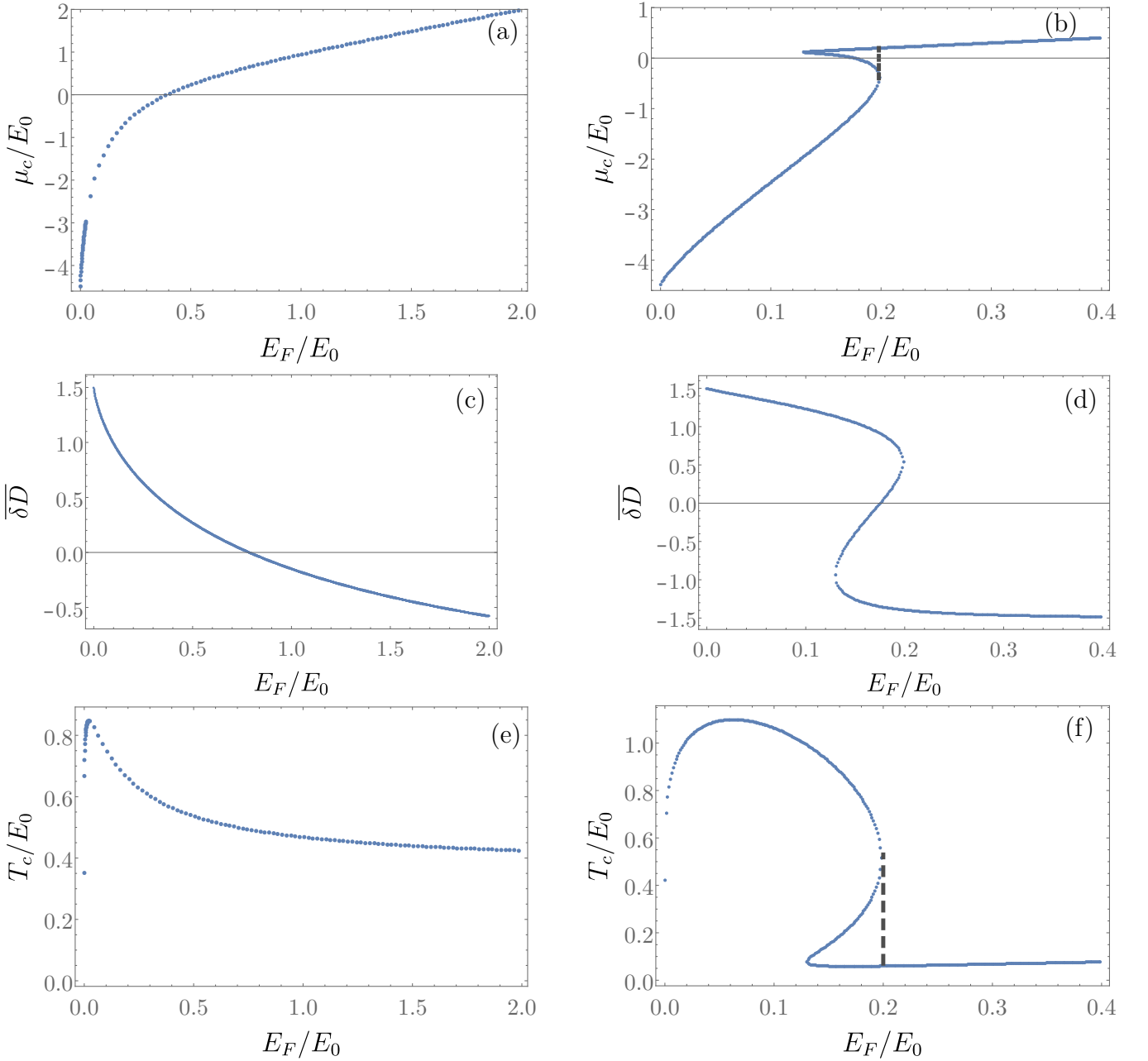
\begin{figure}
    \centering
    \begin{subfigure}[t]{0.48 \linewidth}
    \centering
    \begin{tikzpicture}
    \draw (0,0) node[inner sep=0]{\includegraphics[width = \linewidth]{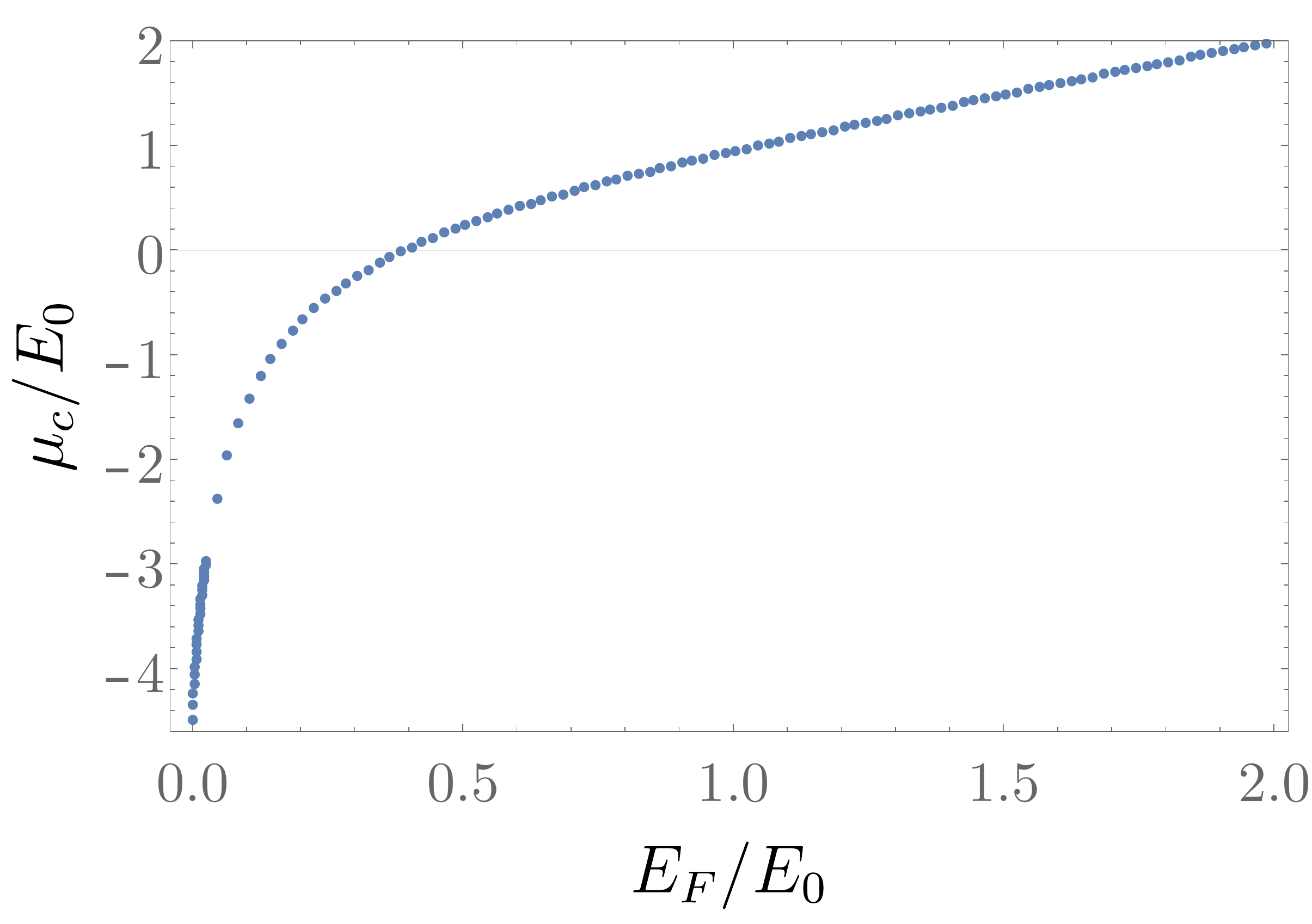}};
    \draw (3.75,2.4) node{\large(a)};
    \end{tikzpicture}
    \end{subfigure}
    \hfill
    \begin{subfigure}[t]{0.48 \linewidth}
    \centering
    \begin{tikzpicture}
    \draw (0,0) node[inner sep=0]{\includegraphics[width = \linewidth]{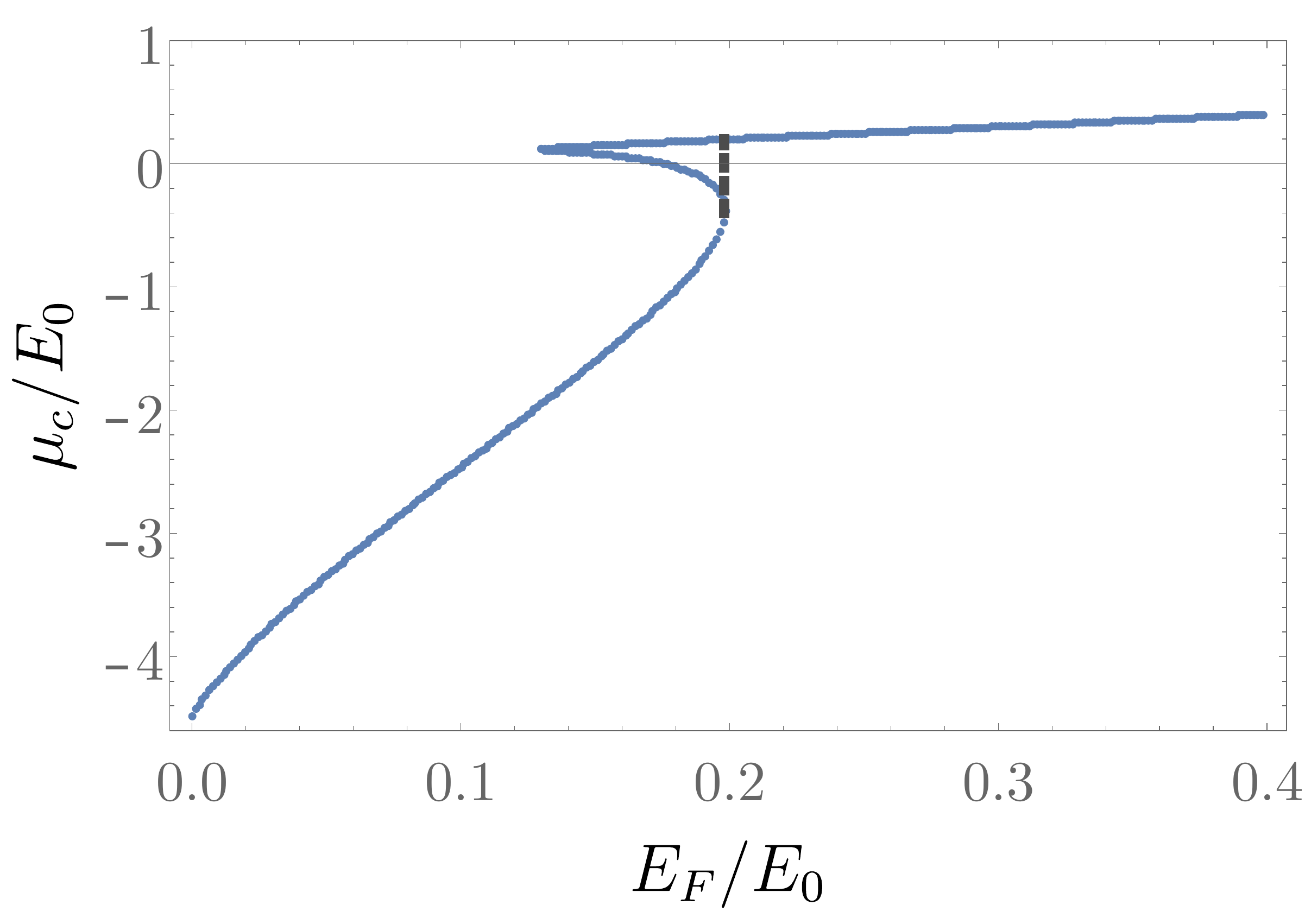}};
    \draw (3.4,2.5) node{\large(b)};
    \end{tikzpicture}
    \end{subfigure}

    \begin{subfigure}[t]{0.45 \linewidth}
    \centering
    \begin{tikzpicture}
    \draw (0,0) node[inner sep=0]{\includegraphics[width = \linewidth]{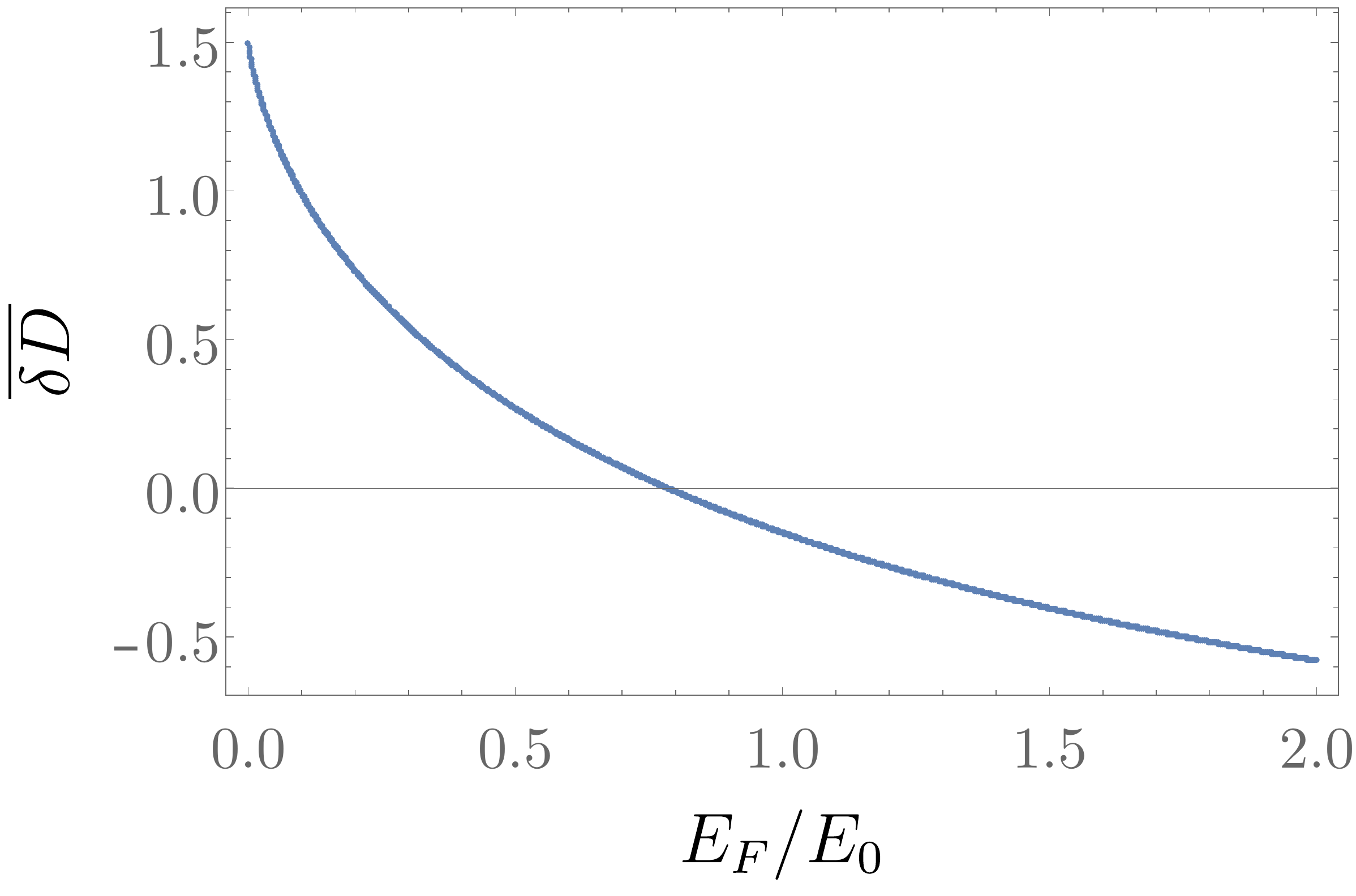}};
    \draw (3.5,2.2) node{\large(c)};
    \end{tikzpicture}
    \end{subfigure}
    \hfill
    \begin{subfigure}[t]{0.45 \linewidth}
    \centering
    \begin{tikzpicture}
    \draw (0,0) node[inner sep=0]{\includegraphics[width = \linewidth]{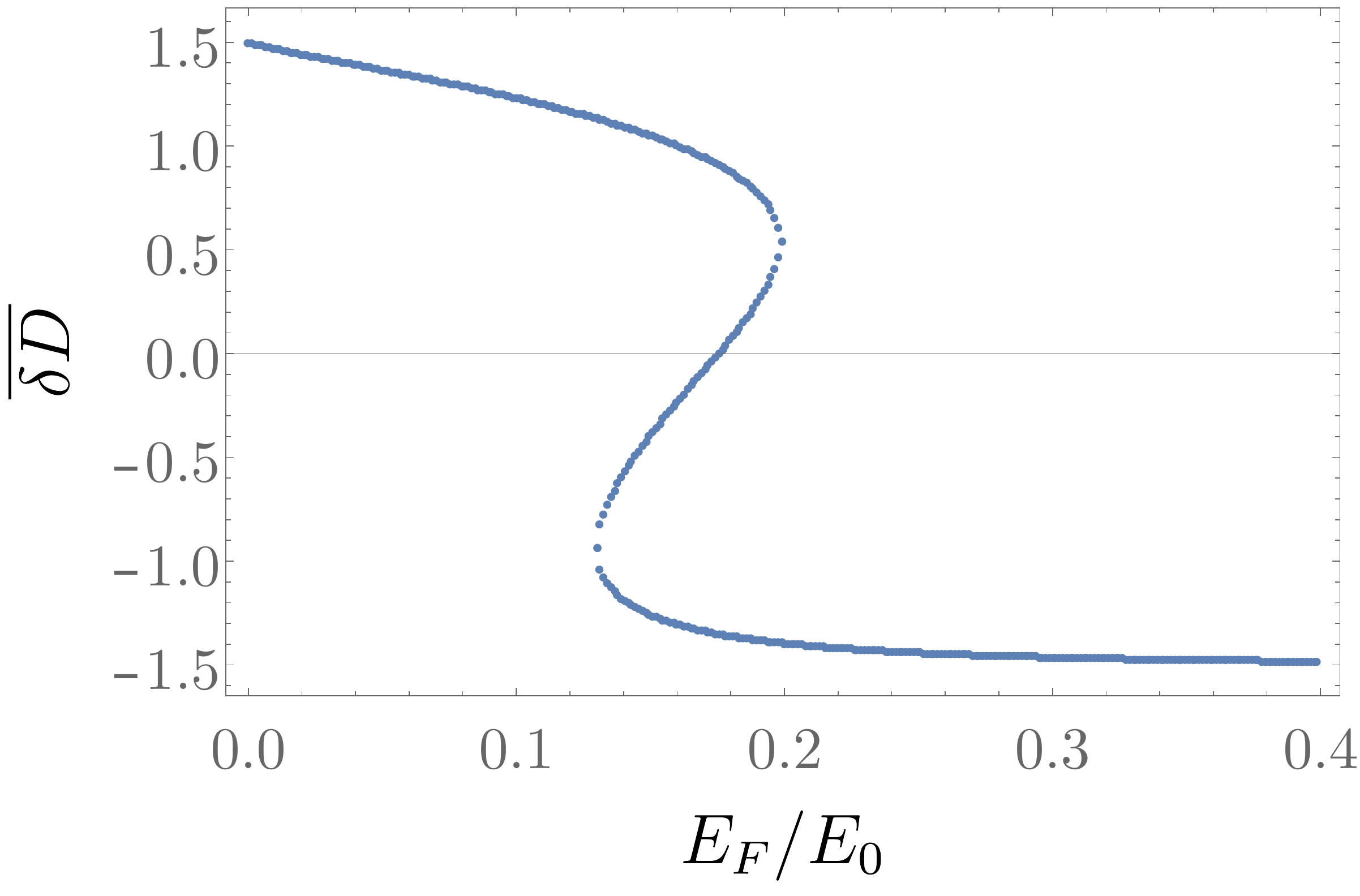}};
    \draw (3.45,2.2) node{\large(d)};
    \end{tikzpicture}
    \end{subfigure}

    \begin{subfigure}[t]{0.48 \linewidth}
    \centering
    \begin{tikzpicture}
    \draw (0,0) node[inner sep=0]{\includegraphics[width = \linewidth]{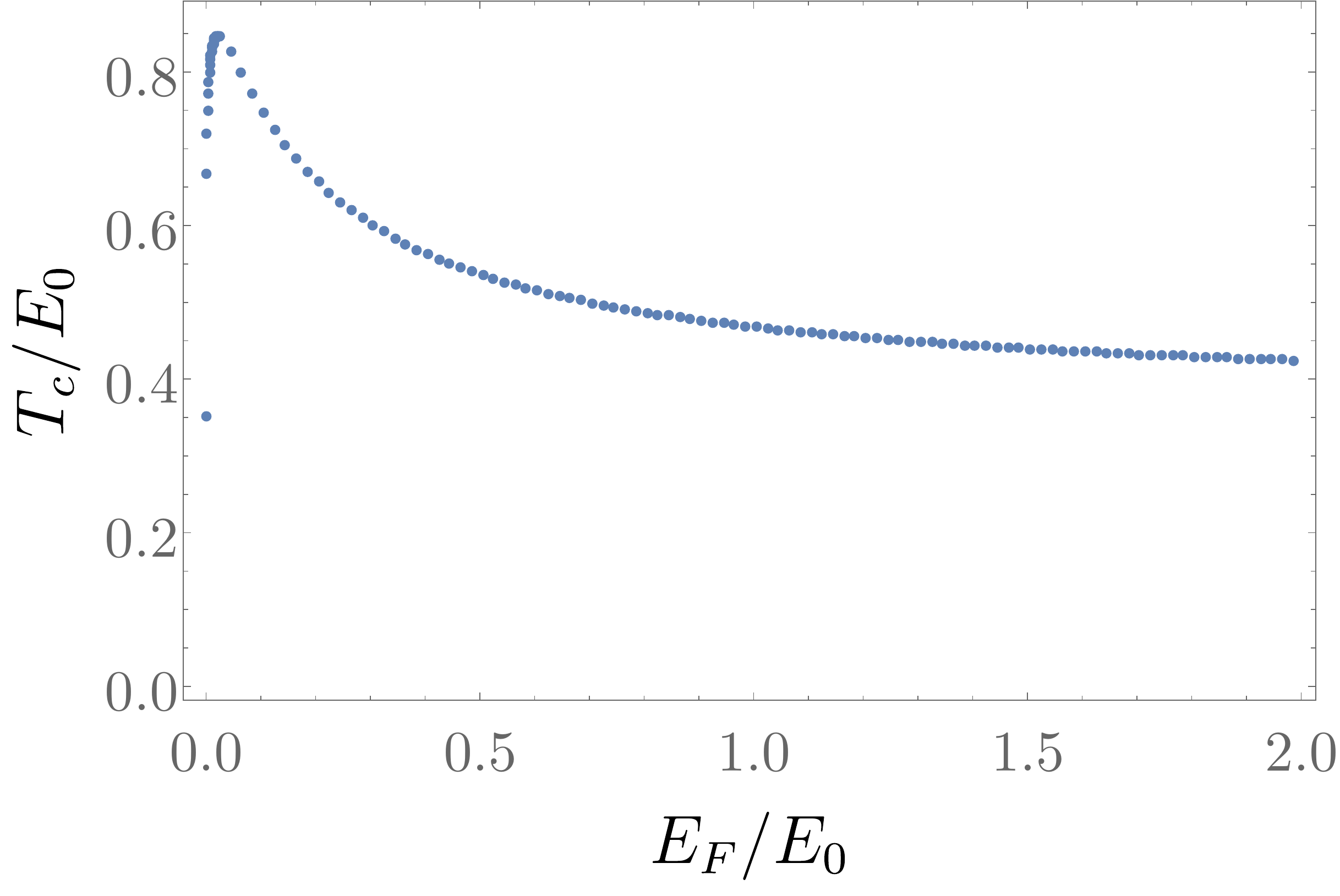}};
    \draw (3.8,2.4) node{\large(e)};
    \end{tikzpicture}
    \end{subfigure}
    \hfill
    \begin{subfigure}[t]{0.48 \linewidth}
    \centering
    \begin{tikzpicture}
    \draw (0,0) node[inner sep=0]{\includegraphics[width = \linewidth]{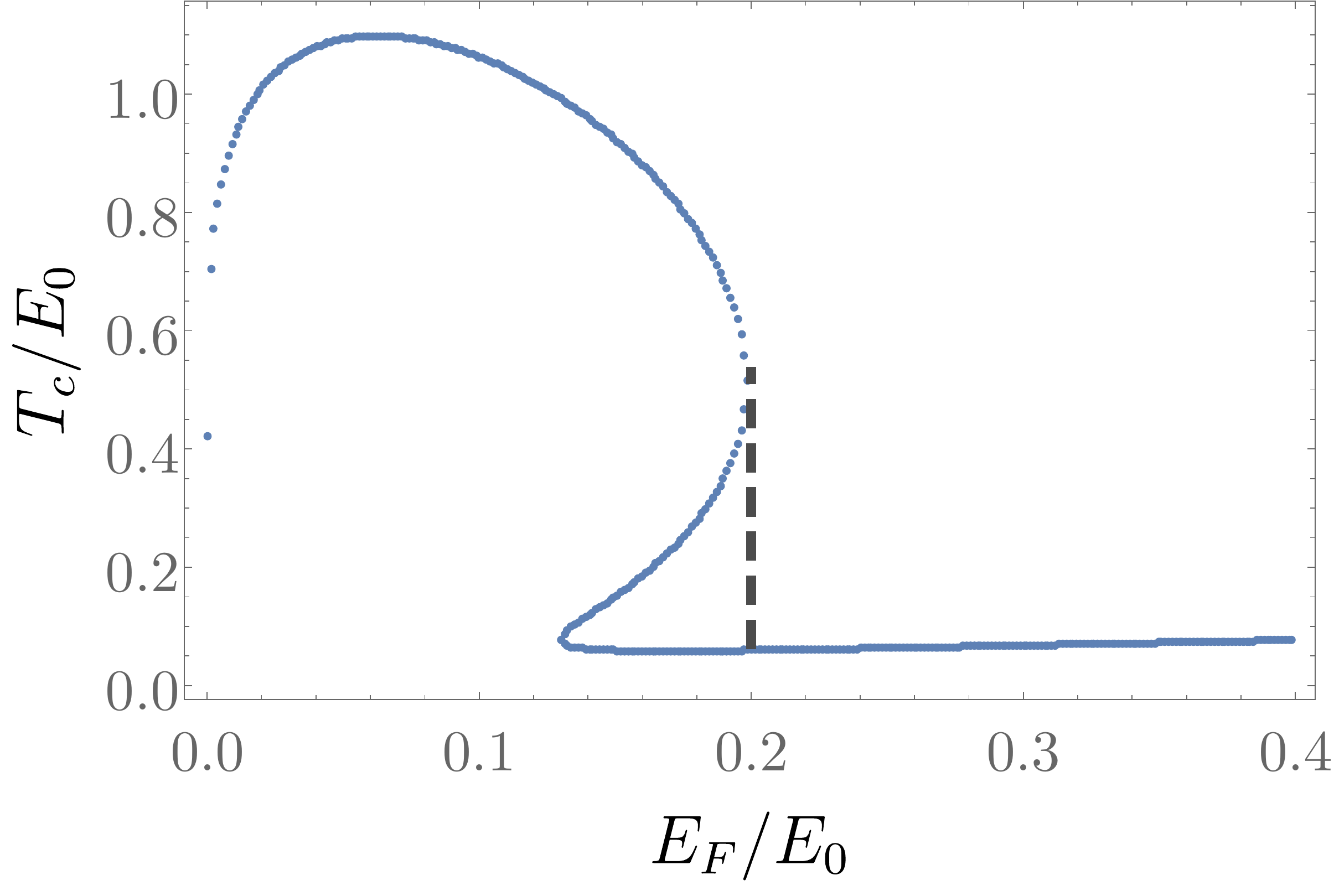}};
    \draw (3.7,2.5) node{\large(f)};
    \end{tikzpicture}
    \end{subfigure}

    \caption{The numerically calculated values of $T_c$,$\mu_c$, and $\overline{\delta D}$ in the region where the KL correction changes sign. The plots on the left are calculated within strict perturbation theory,
    the plots on the right are calculated self-consistently.  Self-consistent calculation yields multi-valued quantities around $E_F \approx 0.2 E_0$, which in practice means that $T_c$ and $\mu_c$ change discontinuously upon variation of $E_F/E_0$
     (dashed lines in the right-hand-side panels for $T_c$ and $\mu_c$).}
    \label{fig:kl}
\end{figure}

\section{Numerical calculation of $T_c$}
For general values of $E_F$, we calculate $T_c$ numerically by simultaneously solving Equations (\ref{eq:chempot2}) and (\ref{eq:fullpairing}). In Fig. (\ref{fig:Tc})  we present the numerical results for $T_c$ in the three regions of $E_F$ ($E_F \gg \omega_D$, $\omega_D \gg E_F \gg E_0$, and $E_0 \gg E_F$) and compare them with our analytic expressions. We see good agreement between analytical and numerical results for all values of $E_F$. This figure summarizes the key results of our work.

In Fig. \ref{fig:D} we present the result of our numerical evaluations of the KL correction $\overline{\delta D}$ over a wide range of $E_F$, obtained by using the numerically obtained $\mu_c(E_F)$ and $T_c(E_F)$.
We see from  Fig. \ref{fig:D} that the KL correction is small for $E_F \gg \omega_D$, in agreement with Migdal's theorem. As $E_F$ is decreased, $\overline{\delta D}$ reaches a sizable finite value close  to $-1.5$ at $E_F \sim 10^3 E_0 \sim 0.1 \omega_D$. Upon further reduction of the particle density, we cross the region where $E_F \sim E_0$. Here, $\overline{\delta D}$ changes sign, and saturates at 1.5 for smaller $E_F/E_0$. This limiting behavior agrees well with our analytical results.

The shaded region in Fig. \ref{fig:D} marks the range near $E_F \sim E_0$, where the result for $\overline{\delta D}$ is more subtle and depends on whether the calculations are done perturbatively or self-consistently.  This also affects the behavior of $T_c$  and $\mu_c$ as functions of $E_F/E_0$. In the perturbative calculation, one computes $\overline{\delta D}$ by using "bare" values of $\mu_c$ and $T_c$, obtained without $\mathcal{O}(\lambda)$ corrections. The bare $\mu_c(E_F)$ is obtained by solving
 Eqs. (\ref{eq:crude}) and (\ref{eq:chempot2}), and is a continuous function of $E_F/E_0$. Additionally, one can show that the bare $\mu_c$ changes sign at $E_F =\frac{2}{\pi}\log(2)e^\gamma E_0 \approx 0.8 E_0$.

Accordingly, $\overline{\delta D}$, computed using the bare $\mu_c(E_F)$ and $T_c(E_F)$, is also a continuous function of $E_F$ and also changes sign at $E_F \approx 0.8 E_0$. We show this perturbative result for $\overline{\delta D}$ in Fig. \ref{fig:kl}c. Combining this perturbative $\lambda \overline{\delta D}$ with other $\mathcal{O}(\lambda)$ corrections, we obtain the result for the renormalized $\mu_c$ and $T_c$, which we present in Fig. \ref{fig:kl}(a,e). We see that while $T_c$ is a continuous function of $E_F/E_0$, it is not monotonic, having a maximum at $E_F \sim 0.04 E_0$.

The problem with the above perturbative calculation is that the bare $\mu_c$ and $T_c$ are used to compute $\overline{\delta D}$, which is highly sensitive to where (at which $E_F/E_0$)  $\mu_c$ changes sign, as well as how $\mu_c$  evolves with $E_F/E_0$. Meanwhile, $\mathcal{O}(\lambda)$ corrections, although nominally small, add factors $\mathcal{O}(1)$ to both $T_c$ and $\mu_c$. This is since both $T_c$ and $\mu_c$ go as $\exp(-2/\lambda)$ for $E_F \lesssim E_0$, and $\mathcal{O}(\lambda)$ corrections to the exponent change both by $\mathcal{O}(1)$).

Therefore the value of $E_F/E_0$ at which $\mu_c$ changes sign, also changes by $\mathcal{O}(1)$. One can see this in Fig. \ref{fig:kl}a, where the "dressed" $\mu_c$ calculated perturbatively changes sign at $E_F \approx 0.4E_0$, rather than $E_F \approx 0.8E_0$ as it did originally. This significantly affects the behavior of $\overline{\delta D}$, which in turn leads to $\mathcal{O}(1)$ corrections to $\mu_c$ and $T_c$. This mutual dependence clearly calls for a fully self-consistent calculation of $T_c$ in the range $E_F \sim E_0$, where $\overline{\delta D}$ rapidly evolves. In other ranges of $E_F/E_0$, where $\overline{\delta D}$ saturates and only weakly varies with $E_F/E_0$, self-consistency is not required.

We show the results of self-consistent calculations of $\mu_c$,  $\overline{\delta D}$, and $T_c$ in the right three panels of Fig.  \ref{fig:kl}. To obtain these results, we treat $\overline{\delta D}$ as a function of $T_c$ and $\mu_c$, and substitute $\overline{\delta D}(\mu_c,T_c)$ into Eq. (\ref{eq:fullpairing}). This equation is then solved self-consistently with Eq. (\ref{eq:chempot2}).
We see from the plots that over some range of $E_F/E_0$, $T_c$ is a multi-valued function of $E_F/E_0$. In practical terms this implies that the superconducting transition temperature (the largest  possible $T_c$ for a given $E_F/E_0$) jumps by a finite amount at $E_F \sim 0.2 E_0$. There is of course a corresponding jump in $\mu_c$ at this $E_F/E_0$. Though subleading corrections to $T_c$ may yield a continuous transition, $T_c$ should change sharply around $E_F/E_0 =0.2$ in either case. Note that the maximum in $T_c$ at smaller values of $E_F$ also emerges in a self-consistent calculation, but is located at a larger $E_F \approx 0.06 E_0$.

\section{Conclusion}
In this paper we derived expressions for the superconducting  $T_c$ with exact prefactors for quasi-2D electrons, with Einstein-phonon-mediated attraction at weak coupling. Previous studies chiefly considered the adiabatic limit  $E_F \gg \omd$. We analyzed $T_c$ in the two other regimes $ E_F \ll E_0$ and $\omd \gg E_F \gg E_0$, where $E_0 = \omd e^{-2/\lambda}$ is the bound state energy for two fermions in a vacuum and $\lambda$ is the dimensionless electron-phonon coupling constant.  In these two regimes the corrections to $T_c$ come from three sources: fermionic self-energy, frequency dependence of the phonon-mediated interaction, and KL renormalization of the pairing interaction by particle-hole excitations. KL corrections are small in $\omd/E_F$ in the adiabatic regime, but become $\mathcal{O}(1)$ in the other two regimes.  We found that the combined renormalization from the three sources reduces $T_c$ from its mean-field value by a factor of almost $10$ in the intermediate regime $E_0 \ll  E_F \ll \omd$, and increases $T_c$ by nearly a factor of $5$ in the regime  $E_F \ll E_0$, which corresponds to very low carrier concentration. We hope that our results will form a starting point for studies of $T_c$ beyond logarithmical accuracy in the physically more relevant case when both electron-electron repulsion and electron-phonon attraction are present.

\begin{acknowledgments}
We are thankful to  M. Christensen, R. Fernandes, M. Gastiasoro, A. Klein, A. Millis, N. Prokofiev, and B. Svistunov for useful discussions. This work was funded by the Department of Energy through the University of Minnesota Center for Quantum Materials, under DE-SC-0016371.
\end{acknowledgments}

\section*{Appendix A: Evaluation of the KL corrections}
Here, we calculate the KL corrections $\delta D(k,q)$ to the interaction. This is a sum of 2 types of diagrams: vertex corrections and exchange corrections, and we will write $\delta D = D_1^\mathrm{vertex}+D_2^\mathrm{vertex}+D^X$. Before calculating these diagrams, we discuss the relevant values of these parameters in the various limits for $E_F$.

As discussed in Section \ref{sec:kl}, we will calculate $\delta D$ at zero external frequency, and with the magnitudes of $\mathbf{k}$ and $\mathbf{q}$ fixed to $\Theta(\mu_c)k_\mu$. Depending on the value of $E_F$, there are essentially 3 limiting regions:

\begin{enumerate}[label=\Alph*)]
    \item $\mu_c < 0$ and $\abs{\mu_c} \gg T_c$: This is where $E_F \ll E_0$.
    \item $\mu_c \approx 0$ and $T_c \gg \abs{\mu_c}$: This region describes the crossover between $E_F \ll E_0$ and $E_0 \ll E_F \ll \omd$.
    \item $\mu_c>0$ and $\mu_c \gg T_c$: This includes the regions $E_0 \ll E_F \ll \omd$ and $E_F \gg \omd$.
\end{enumerate}

In both regions A and C, we have $\abs{\mu_c} \gg T_c$. Since $T_c$ is therefore smallest energy scale in the calculation, we may simply evaluate these diagrams at $T=0$ as an approximation. We also fix the external frequencies equal to zero in both region A and C.

Regarding region B where $\mu_c \approx 0$, we cannot evaluate these diagrams at $T=0$, since $T_c$ is not the smallest energy scale in the problem. However, we will still calculate these diagrams at zero external frequency and momenta. The validity of setting the external frequency and momenta to zero will be discussed below.

\section{Vertex Corrections}
Let us first consider the vertex corrections, denoted $D_1^{\mathrm{vertex}}$ and $D_2^{\mathrm{vertex}}$. One can verify that these two corrections will end up being equal, so we will calculate $D_1^\mathrm{vertex}$ and take $D^\mathrm{vertex} = 2D_1^\mathrm{vertex}$. Referring to Figure \ref{fig:KL0}, we write the expression for $D^\mathrm{vertex}$ below, where we have used $D_0(\om) = \omd^2/(\om^2+\omd^2)$, $G(k) = (i\om-\varepsilon_k)^{-1}$, and $\varepsilon_k = k^2/2m-\mu_c$:

\begin{equation}
    \lambda D^\mathrm{vertex} = 2g T_c D_0(\omega_p) \sum_{\Om} \int
    \frac{d^2l}{(2\pi)^2} \frac{1}{i\Om+i\omega_q-\varepsilon_l}\frac{1}{i\Om+i\omega_q+i\omega_p-\varepsilon_{l+p}} D_0(\Om).
\end{equation}

In the above expression, we have defined $\mathbf{p}=\mathbf{k}-\mathbf{q}$ and $\omega_p=\omega_k-\omega_q$, and $\Om$ is a bosonic Matsubara frequency. Since $\omega_p = \omega_k-\omega_q $ is on the order of $T_c \ll \omd$, we may replace $D_0(\omega_p)=1$. Using partial fractions, we find

\begin{equation}
    \lambda D^\mathrm{vertex} = 2g T_c \sum_{\Om} \int
    \frac{d^2l}{(2\pi)^2}
    \frac{1}{\varepsilon_{l+p}-\varepsilon_l-i\omega_p}
    \big(\frac{1}{i\Om+i\omega_q+i\omega_p-\varepsilon_{l+p}}-\frac{1}{i\Om+i\omega_q-\varepsilon_l }\big)
    D_0(\Om)
\end{equation}

Calculating the Matsubara sum, and setting $n_B(\omd) = 0$ (since $T_c \ll \omd$), we have

\begin{multline} \label{eq:gen_a}
    \lambda D^\mathrm{vertex} = g \omd \int
    \frac{d^2l}{(2\pi)^2}
    \frac{1}{\varepsilon_{l+p}-\varepsilon_l-i\omega_p}
    \big(
    \frac{n_F(\varepsilon_{l+p})}{i\omega_q+i\omega_p-\varepsilon_{l+p}+\omd}\\
    -
    \frac{1-n_F(\varepsilon_{l+p})}{\omd+\varepsilon_{l+p}-i\omega_q-i\omega_p}
    -\frac{n_F(\varepsilon_l)}{i\omega_q-\varepsilon_l+\omd}
    +
    \frac{1-n_F(\varepsilon_l)}{\omd+\varepsilon_l-i\omega_q}
    \big).
\end{multline}

Now, we may simplify this our 3 different limits.

\subsection{Region A}
Let us begin with region A, where $\abs{\mu_c} \gg T_c$ and $\mu_c < 0$. All Fermi functions are effectively zero in this region, and we have after some algebra

\begin{equation} \label{eq:regA}
    \lambda D_A^\mathrm{vertex}  = g \omd \int \frac{d^2l}{(2\pi)^2}
    \frac{1}{\omd+\varepsilon_l-i\omega_q}\frac{1}{\omd+\varepsilon_{l+p}-i\omega_q-i\omega_p}
\end{equation}

As discussed above, in region A, we set all external frequencies and momenta to zero. This expression then becomes

\begin{align} \label{eq:reg}
    \lambda  D_A^\mathrm{vertex}
    &=
    g N_0 \omd \int_{-\mu_c}^{\infty} d\varepsilon
    \frac{1}{(\omd+\varepsilon)^2}\\ \label{eq:vertexA}
    &=
    \lambda \frac{\omd}{\omd-\mu_c}.
\end{align}

Since $\mu_c \ll \omd$ in region A, we have $D_A^\mathrm{vertex} \approx 1$. As alluded to in the main text, this KL correction is nonzero in region A (where $\mu_c < 0$). From this calculation, we see that this is due to the dynamical nature of our interaction (more precisely, the presence of a pole in our bosonic propagator.)

\subsection{Region B}
In this region, we must now include the terms with Fermi functions. For reasons that will become clear below, we will refer to this as the singular part of $D^\mathrm{vertex}$. Regarding the terms without Fermi functions, we may simply take our above result from region A, since we are still working at zero external momenta and frequencies. We will refer to this expression as the regular part of $D^\mathrm{vertex}$. Focusing on the singular part of $D^\mathrm{vertex}$, we have

\begin{multline} \label{eq:sing_exact}
    \lambda D^\mathrm{vertex}_\mathrm{sing} = g \omd \int \frac{d^2l}{(2\pi)^2}
    \frac{1}{\varepsilon_{l+p}-\varepsilon_l-i\omega_p}
    \big(
    \frac{n_F(\varepsilon_{l+p})}{i\omega_q+i\omega_p-\varepsilon_{l+p}+\omd}
    +
    \frac{n_F(\varepsilon_{l+p})}{\omd+\varepsilon_{l+p}-i\omega_q-i\omega_p}\\
    -
    \frac{n_F(\varepsilon_l)}{i\omega_q-\varepsilon_l+\omd}
    -
    \frac{n_F(\varepsilon_l)}{\omd+\varepsilon_l-i\omega_q}
    \big).
\end{multline}

Since $\omd$ is the largest energy scale in region B, we simply replace the denominators of all Fermi functions by $\omd$, obtaining

\begin{align} \label{eq:sing}
    \lambda D^\mathrm{vertex}_\mathrm{sing} \approx 2g \int \frac{d^2l}{(2\pi)^2}
    \frac{n_F(\varepsilon_{l+p})-n_F(\varepsilon_l)}{\varepsilon_{l+p}-\varepsilon_l-i\omega_p}.
\end{align}

Working in the static limit, and taking $p \rightarrow 0$, we find

\begin{align}
    \lambda D_\mathrm{sing}^\mathrm{vertex}(p=0) &= 2g \int \frac{d^2l}{(2\pi)^2}
    \frac{dn_F(\varepsilon_l)}{d\varepsilon_l}\\
     &=
     2g N_0 \int_{-\mu_c}^\infty d\varepsilon \frac{dn_F(\varepsilon)}{d\varepsilon} \\
     &= -2\lambda n_F(-\mu_c),
\end{align}

so we have $D_\mathrm{sing}^\mathrm{vertex}(p=0) =-2n_F(-\mu_c) $. Note that this Fermi function leads to a step-like jump as we transition from region A to region C through region B. This is why we refer to it as singular. In contrast, the other part of this vertex correction is essentially 1 across the transition, which is why we called it the regular part of $D^\mathrm{vertex}$. Putting $D^\mathrm{vertex}_\mathrm{reg}$ and $D^\mathrm{vertex}_\mathrm{sing}$ together, we find $D_B^\mathrm{vertex}(p=0) = -\tanh \frac{\mu_c}{2T_c}$.

This is what we use in the numerical calculations of $T_c$. As we will see below, this expression which was evaluated at $p=0$ overestimates the effect of the singular piece in the crossover region. However, it has the correct qualitative behavior, i.e. the vertex corrections smoothly decrease from 1 to -1 connecting the limiting behaviors of both region A and region C.

With this, we now turn to the complications discussed above, that we cannot naively evaluate this diagram at $p=\omega_p=0$. Instead we must consider momenta $q$ and $k$ such that $\varepsilon_q$ and $\varepsilon_k$ are on the order of $T_c$. We will see that in this crossover region where $\abs{\mu_c} \ll T_c$, the $D_\mathrm{sing}^\mathrm{vertex}(p)$ dies quickly with increasing $p$. This invalidates the assumption made in the main text, that the KL diagram is relatively constant over the region of $\mathbf{q}$ which contribute significantly to the integral. In fact, the quick decay of $D_\mathrm{sing}^\mathrm{vertex}(p)$ with $p$ destroys the logarithmical singularity in the second term of Eq. \ref{eq:fullgap}. Therefore, the effect on $T_c$ in region B due to the vertex correction is not due to the singular piece, but the regular piece.

To show this, let us rewrite $D_\mathrm{sing}^\mathrm{vertex}(p)$ and take $\mu_c = 0$ for convenience. Using Equation \ref{eq:sing}, we have

\begin{align}
    \lambda D^\mathrm{vertex}_\mathrm{sing}(p)
    &=
    2g \int \frac{d^2l}{(2\pi)^2}
    \frac{n_F(\varepsilon_{l+p})-n_F(\varepsilon_l)}{\varepsilon_{l+p}-\varepsilon_l-i\omega_p}\\
    &=
    2 g \int \frac{d^2l}{(2\pi)^2}
    \big(
    \frac{n_F(\varepsilon_l)}{\varepsilon_l-\varepsilon_{l+p}-i\omega_p}
    -
    \frac{n_F(\varepsilon_l)}{\varepsilon_{l+p}-\varepsilon_l-i\omega_p}
    \big)\\
    &=
    -2 g P\int \frac{d^2l}{(2\pi)^2}
    \frac{n_F(\varepsilon_l)}{\varepsilon_{l+p}-\varepsilon_l}\\
\end{align}

To obtain the last equality, we took $\omega_p = 0$ for simplicity. Doing the angular integration, we find

\begin{align}
    D^\mathrm{vertex}_\mathrm{sing}(p)
    &=
    -\frac{2}{p} \int_0^{p/2} \frac{dl l n_F(\varepsilon_l)}{\sqrt{(p/2)^2-l^2}}\\
    &\approx
    -2n_F(\varepsilon(p/2)).
\end{align}

To obtain the final expression, we first notice that the integrand peaks when $l=p/2$. Therefore, as a crude approximation, we pull $n_F(\varepsilon_l)$ outside of the integral, evaluated at $l=p/2$, and do the remaining integral. As asserted above, this expression decays quickly with $p$. One may verify that this indeed leads to the disappearance of the logarithmical singularity upon insertion into Eq. \ref{eq:fullgap}.

\subsection{Region C}
As in region A, we have $\abs{\mu_c} \gg T_c$ and we can again evaluate this expression at $T=0$. However, both the regular and singular pieces of the vertex correction now contribute since $\mu_c>0$ (we cannot set the Fermi functions to zero). However, we can use our $T=0$ approximation to replace the Fermi function $n_F(x)$ with $\Theta(-x)$. Using this and $1-\Theta(-x) = \Theta(x)$, we find

\begin{equation}
    \lambda D_C^\mathrm{vertex} = g \omd P \int
    \frac{d^2l}{(2\pi)^2}
    \frac{1}{\varepsilon_{l+p}-\varepsilon_l}
    \big(
    \frac{\Theta(-\varepsilon_{l+p})}{-\varepsilon_{l+p}+\omd}
    -
    \frac{\Theta(\varepsilon_{l+p})}{\omd+\varepsilon_{l+p}}
    -\frac{\Theta(-\varepsilon_l)}{-\varepsilon_l+\omd}
    +
    \frac{\Theta(\varepsilon_l)}{\omd+\varepsilon_l}
    \big).
\end{equation}

We rewrite the step functions in terms of $\sgn$ functions, obtaining

\begin{align}
    \lambda D_C^\mathrm{vertex}
    &=
    g \omd P \int
    \frac{1}{\varepsilon_{l+p}-\varepsilon_l}
    \big(
    -\frac{\sgn(\varepsilon_{l+p})}{\omd+\abs{\varepsilon_{l+p}}}
    +
    \frac{\sgn(\varepsilon_l)}{\omd+\abs{\varepsilon_l}}
    \big)\\
    &=
    2g \omd P \int \frac{l dl}{2\pi} \frac{\sgn(\varepsilon_l)}{\omd+\abs{\varepsilon_l}}
    \int \frac{d\theta}{2\pi} \frac{1}{p^2/2m+ l p \cos(\theta)/m}.
\end{align}

Doing the angular integration and canceling a factor of $\lambda = g N_0$, we find

\begin{equation} \label{eq:cvertex}
    D_C^\mathrm{vertex} = \frac{2\omd}{p} \int_0^{p/2}
    \frac{l dl}{\sqrt{(p/2)^2-l^2}} \frac{\sgn(\varepsilon_l)}{\omd+\abs{\varepsilon_l}}.
\end{equation}

Since in region C, we set the external momenta $q$ and $k$ equal to $k_\mu$, we have $p = \abs{\mathbf{q}-\mathbf{k}}=2k_\mu \sin\theta/2$, where $\theta$ is the angle between $\mathbf{q}$ and $\mathbf{k}$. Therefore, the upper limit of this integral is $p/2=k_\mu \sin\theta/2 < k_\mu$. Since $\varepsilon_l = l^2/2m-\mu_c$, if $l < k_\mu$ (as in the above integral), we have $\varepsilon_l < 0$. Therefore, we set $\sgn(\varepsilon_l)=-1 $ and $\abs{\varepsilon_l}=\mu_c-l^2/2m$, obtaining

\begin{equation}
    D_C^\mathrm{vertex} = -\frac{2\omd}{p} \int_0^{p/2}
    \frac{l dl}{\sqrt{(p/2)^2-l^2}} \frac{1}{\omd+\mu_c-l^2/2m}.
\end{equation}

We now rescale $l=k_\mu x$ and $\bmuc=\mu_c/\omd$, plug in $p=2k_\mu \sin \theta/2$, and average over all $\theta$ to obtain

\begin{equation}
    \overline{D_C^\mathrm{vertex}(\bmuc)}
    =
    -\frac{1}{\pi}\int_0^\pi d\theta \frac{1}{\sin(\theta/2)}
    \int_0^{\sin(\theta/2)} \frac{x dx}{\sqrt{\sin^2(\theta/2)-x^2}} \frac{1}{1+\bmuc (1-x^2)}.
\end{equation}

This is the expression we use to numerically calculate the vertex corrections for any $\mu_c$ in region C. We can simplify this expression analytically in the limit of large and intermediate density ($E_F \gg \omd$ and $\omd \gg E_F \gg E_0$), obtaining $\overline{D_C^\mathrm{vertex}} = 0$ and $\overline{D_C^\mathrm{vertex}} = -1$ respectively. Note that the vertex corrections go to zero in the limit of small $\omd/E_F$ in accordance with Migdal's theorem.

\section{Exchange Diagram}
We now move on to the exchange diagram, which we will denote $D^X$. We have from Figure \ref{fig:KL0}

\begin{equation}
    \lambda D^X = g T \sum_{\Om} \int \frac{d^2l}{(2\pi)^2}
    G_0(l)G_0(l+p) D_0(\Om-\omega_q) D_0(\Om-\omega_k),
\end{equation}

where we have redefined $\omega_p=-(\omega_q+\omega_k)$, $\mathbf{p} = -(\mathbf{q}+\mathbf{k})$, and $\Om$ is a fermionic Matsubara frequency. If we let $\Om \rightarrow \Om-\omega_q$, we have instead

\begin{equation}
    \lambda D^X = g T \sum_{\Om} \int \frac{d^2l}{(2\pi)^2}
    \frac{1}{i\Om+i\omega_q-\varepsilon_l}\frac{1}{i\Om+i\omega_q+i\omega_p-\varepsilon_{l+p}} D_0(\Om) D_0(\Om+\omega_q-\omega_k),
\end{equation}

where the redefined $\Om$ is now bosonic. Since $\omega_k$ and $\omega_q$ are on the order of $T \ll \omd$, we may approximate this sum with

\begin{equation}
    \lambda D^X = g T \sum_{\Om} \int \frac{d^2l}{(2\pi)^2}
    \frac{1}{i\Om+i\omega_q-\varepsilon_l}\frac{1}{i\Om+i\omega_q+i\omega_p-\varepsilon_{l+p}} D_0(\Om)^2.
\end{equation}

Note that the expressions for $D^\mathrm{vertex}$ and $D^X$ are identical except for a difference of 2 in the prefactor and the fact that $D_0(\Om)$ is squared in the exchange diagram. We will exploit this similarity to obtain expressions for $D^X$ in all three regions from our previous work. Using the definition of $D_0(\Om)$, one may verify

\begin{equation}
    \frac{d}{d\omd^2} \frac{D_0(\Om)}{\omd^2} = -\frac{D_0(\Om)^2}{\omd^4}.
\end{equation}

We therefore have

\begin{equation} \label{eq:derive}
    D^X(p,\omega_p) = -\frac{\omd^4}{2} \frac{d}{d\omd^2}( \frac{1}{\omd^2} \dvertex(p,\omega_p)).
\end{equation}

We have explicitly written the dependence on $\omega_p$ and $p$ to emphasize that this identity is true only before we write how $p$ depends on $k$ and $q$. This is because in the case of the vertex corrections above, $p = k-q$, while for the exchange corrections, $p=-q-k$. We now use this derivative formula to calculate $D^X$ in our three limits.

\subsection{Region A}
In region A, we have $p = -q-k = 0$. Applying our derivative formula (Equation \ref{eq:derive}) to our previous result in region A, $\dvertex=1/(1-\bmuc)$, we obtain

\begin{equation}
    D_A^X = \frac{1}{4}\frac{2-\bmuc}{(1-\bmuc)^2} \approx \frac{1}{2}.
\end{equation}

\subsection{Region B}
In region B, we also have $p=0$. Applying our derivative formula to our previous equation in Region B, we obtain

\begin{equation}
    D_B^X = -\frac{1}{2} \tanh \frac{\mu_c}{2T_c}.
\end{equation}

As before, this is a sum of regular and singular parts, with the singular piece switching on across $\mu_c=0$. As before, this expression overestimates the effect of the singular piece, which does not affect $T_c$ until $\abs{\mu_c}$ exceeds $T_c$.

\subsection{Region C}
Applying the derivative formula to Equation \ref{eq:cvertex}, we find

\begin{align}
    D_C^X(k,q)
    &=
    \frac{1}{2p} \int_0^{p/2}l dl \frac{\sgn{\varepsilon_l}}{\sqrt{(p/2)^2-l^2}}
    \frac{2+|\bar{\varepsilon_l}|}{(1+|\bar{\varepsilon_l}|)^2},
\end{align}

where we have defined $\overline{\varepsilon_l}=\varepsilon_l/\omd$. Since we have $k=q=k_\mu$ in region C, we have $p=\abs{\mathbf{k}+\mathbf{q}} = 2k_\mu \cos \theta/2$. We can now simplify this expression as before to obtain

\begin{equation}
    \overline{D_C^X(\bmuc)}
    =
    -\frac{1}{4\pi}\int_0^\pi \frac{d\theta}{\cos(\theta/2)}\int_0^{\cos(\theta/2)}
    dx \frac{x}{\sqrt{\cos^2(\theta/2)-x^2}}\frac{2+\bmuc (1-x^2)}{(1+\bmuc(1-x^2))^2}
\end{equation}

This is the expression we use to numerically calculate the exchange contribution for any $\bmuc$ in region C. As before, we can simplify this expression analytically when $\mu_c \ll \omd$ (corresponding to $E_0 \ll E_F \ll \omd$) and $\mu_c \gg \omd$ (corresponding to $E_F \gg \omd$), obtaining -1/2 and 0 respectively.

\subsubsection{Total KL Contribution}
The total correction to the interaction, $\delta D$, is found by summing the contribution from the vertex corrections and the exchange diagram, $\delta D=\dvertex+\dex$. We can now numerically calculate KL contribution at any $\mu_c$, given the region (A, B, or C) in which $\mu_c$ exists. Additionally, though we do not have an analytic expression that holds for general $\mu_c$, we have obtained simple results for this correction in our general limits of $E_F$, which are summarized in Table \ref{tab:KLtab}. The total correction as a function of $E_F/E_0$ as been plotted in Figure \ref{fig:D}.

\begin{table}[]
    \centering
    \begin{tabular}{c|c|c|c}
    & $\overline{\dvertex}$ & $\overline{\dex}$ & $\overline{\delta D}$  \\
    \hline
    $E_F \ll E_0$   & 1 & 1/2 & 3/2 \\
    $E_0 \ll E_F \ll \omd$ & -1  & -1/2  & -3/2  \\
    $\omd \ll E_F$ & 0  & 0    & 0    \\
    \hline
    \hline
    \end{tabular}
    \caption{The summary of the analytic results of this paper regarding the KL corrections. These directly affect the prefactor of $T_c$ in each region for $E_F$.}
    \label{tab:KLtab}
\end{table}

\section*{Appendix B: Details of numerical calculations}
For our numerical calculation of $T_c$, we start from the full equation for the pairing vertex, Eq. \ref{eq:pairing_vertex}, which we rewrite below for convenience.

\begin{align}
    \Phi(\om,\mathbf{k})
    &=
    -T \sum_{\Om} \int \frac{d^2q}{(2\pi)^2}G(\Omega_m,q) G(-\Omega_m, -q) V_\mathrm{eff} (\om,\mathbf{k};\Om,\mathbf{q})\Phi(\Om,\mathbf{q}).
\end{align}

In the main text, we discussed all three effects on $T_c$ separately, and added their contributions at the end, which is valid at weak coupling. Doing so, our expression for the pairing vertex becomes

\begin{equation} \label{eq:numerical}
    \frac{T_c}{N_0}\sum_{\Om} \int \frac{d^2q}{(2\pi)^2} \frac{1}{Z^2\Om^2+\varepsilon_q^2}
    \frac{\omd^4}{(\omd^2+\Om^2)^2}
    = \frac{Z}{\lambda} - \overline{\delta D}.
\end{equation}

Integrating over momentum and doing one of the frequency sums, this equation becomes

\begin{equation} \label{eq:numerical}
    \frac{1}{2}\log(\frac{2e^\gamma}{\sqrt{e}\pi \bar{T}_c})+\bar{T}_c\sum_{\bOm} \arctan(\frac{\bmuc}{\bOm})\frac{1}{(1+\bOm^2)^2}\frac{1}{\abs{\bOm}}
    = \frac{Z(\bmuc)}{\lambda} - \overline{\delta D}(\bmuc),
\end{equation}

where we have redefined all variables to be relative to $\omd$, and emphasized the fact that Z and $\overline{\delta D}$ are functions of $\bmuc$. We now rewrite the second term on the left-hand-side as follows:
\begin{equation} \label{simpsum}
    \bar{T}_c\sum_{\bOm} \arctan(\frac{\bmuc}{\bOm})\frac{1}{(1+\bOm^2)^2}\frac{1}{\abs{\bOm}}
    =
    \int_0^{\bmuc} dx \bigg(
    \frac{-1}{x^2-1}\frac{1}{8\bar{T}_c}\sech^2\frac{1}{2\bar{T}_c}
    +\frac{x^2-3}{4(x^2-1)^2}\tanh{\frac{1}{2\bar{T}_c}}
    +\frac{1}{2x(x^2-1)^2}\tanh{\frac{x}{2\bar{T}_c}}
    \bigg)
\end{equation}
This integral is more convenient than the original sum for computational purposes. To derive this expression, recall that for any function $f(\bmuc)$, $f(\bmuc)=\int_0^{\bmuc} \frac{df(x)}{dx}dx$, assuming $f(0)=0$. This is what we did above, where after taking the derivative, the sum has been evaluated explicitly. This simplified version of the equation for the pairing vertex is then solved simultaneously with $\bmuc = \bar{T}_c \log(\exp(\bar{E}_F/\bar{T}_c)-1)$ for a given $E_F$.

%DP begin changes
Though this can in principle be done for any $E_F$, we only use the above equation for $E_F >5E_0$, and instead solve a simplified equation for $E_F < 5E_0$, where $\abs{\mu_c} \ll \omd$. For $E_F > 5E_0$, we also use the Kohn-Luttinger expression calculated in Region C, from the above Appendix. The threshold of $E_F = 5E_0$ is of course artificial. We only require $\abs{\mu_c} \ll \omd$ for our simplified equation to apply, and we find that at $E_F = 5E_0$, $\mu_c/\omd = 2 \times 10^{-4} \ll 1$ (using $\lambda = 0.2$).

In the region where $\abs{\mu_c} \ll \omd$, the above sum can be simplified to

\begin{equation}
    \bar{T}_c\sum_{\bOm} \arctan(\frac{\bmuc}{\bOm})\frac{1}{\bOm}
    =
    \int_0^{\bmuc} dx
    \frac{1}{2x}\tanh{\frac{x}{2\bar{T}_c}}.
\end{equation}

Additionally, for $\abs{\mu_c} \ll \omd$, we may set $Z = 1+\lambda/2$. The resulting equation for $T_c$ can be written as

\begin{equation}
    \log(\frac{2e^{\gamma-3/2}}{\pi \widetilde{T_c}})-3\tanh\frac{\widetilde{\mu}_c}{2\widetilde{T_c}}+\int_0^{\widetilde{\mu}_c}\frac{\tanh\frac{x}{2\widetilde{T_c}}}{x} = 0,
\end{equation}

where all quantities with tildes have been expressed in terms of $E_0$. Note that we have used the expression for the Kohn-Luttinger correction which applies only in the crossover region (region B), where $\mu_c \ll T_c$. It is in fact unnecessary to use the expression calculated in region A ($E_F \ll E_0$), since the expression in region B smoothly saturates to the constant value calculated in region A. This is the computationally more convenient equation we solve(both self-consistently and non-self-consistently), along with the equation for the chemical potential for $T_c$ and $\mu_c$ for $E_F \leq 5E_0$. All numerical results are obtained using $\lambda = 0.2$.
%DP end changes

\section*{Appendix C: Assumption of Bandwidth}
Throughout this paper, we have worked in the infinite bandwidth limit, i.e. the bandwidth $\Lambda$ is much larger than $E_F$ and $\omd$. In general, the effect of the Kohn-Luttinger corrections will depend on the ratio $\Lambda/\omd$. To illustrate this point, we will evaluate the vertex correction $D^\mathrm{vertex}$ at finite bandwidth. For simplicity, we will evaluate this correction at $\mu_c < 0$, for $\abs{\mu_c} \gg T_c$ (see Region A.) Referring to our calculations in Appendix A, we may simply take Eq. \ref{eq:vertexA} and replace the upper limit by $\Lambda$. The new result at finite bandwidth is then

\begin{equation}
    D_A^\mathrm{vertex} = \frac{\omd}{\omd-\mu_c}-\frac{\omd}{\omd-\Lambda}.
\end{equation}

If we work in the limit where $\abs{\mu_c} \ll \omd$ and $\Lambda$, the expression simplifies to $D_A^\mathrm{vertex} = \Lambda/(\Lambda-\omd)$. Note that if we work in the limit where $\Lambda \gg \omd$, we have $D_A^\mathrm{vertex} = 1$, and we retrieve the result obtained in appendix A. However, if we work in the opposite limit where $ \omd \gg \Lambda$, we obtain $D_A^\mathrm{vertex} \approx - \Lambda/\omd \rightarrow 0$. This agrees with previous work on Kohn-Luttinger corrections that considered static interactions \cite{Chubukov2016,pisani2018entanglement}, which found that the Kohn-Luttinger corrections disappear in the low-density limit ($\mu_c < 0$.)

\bibliographystyle{apsrev}
\bibliography{references}

\end{document}